\documentclass[letterpaper, 11pt, final, onecolumn]{IEEEtran}
\usepackage[margin=1.0in]{geometry}
\usepackage{times}
\usepackage[pdftex]{graphicx}
\usepackage{subfigure}
\usepackage{xcolor}
\usepackage{rotating}
\usepackage{rotfloat}
\usepackage{amsmath}
\usepackage{amssymb}
\usepackage{algorithm}
\usepackage{algorithmic}
\usepackage{gensymb}
\usepackage{textcomp}
\usepackage{balance}
\usepackage{placeins}
\usepackage{cite}
\usepackage{url}
\usepackage{booktabs}

\begin{document}

\title{SaYoPillow: A Blockchain-Enabled, Privacy-Assured Framework for Stress Detection, Prediction and Control Considering Sleeping Habits in the IoMT}


\author
{

\begin{tabular}{cc}
	\\\\
Laavanya Rachakonda & Anand K. Bapatla \\
	Dept. of Computer Science and Engineering & Dept. of Computer Science and Engineering \\
University of North Texas, USA. & University of North Texas, USA.\\
	Email: RachakondaLaavanya@my.unt.edu & Email: AnandKumarBapatla@unt.edu
	\\\\
Saraju P. Mohanty &  Elias Kougianos  \\
	Dept. of Computer Science and Engineering & Dept. of Computer Science and Engineering \\
University of North Texas, USA. & University of North Texas, USA.\\
	Email: saraju.mohanty@unt.edu & Email: elias.kougianos@unt.edu
	\\
\end{tabular}

}

\maketitle

\begin{abstract}
Considering today's lifestyle, people just sleep forgetting the benefits it provides to the human body. The reasons for not having a productive sleep could be many. \textbf{S}m\textbf{a}rt-\textbf{Yo}ga Pillow (SaYoPillow) is envisioned as a device that may help in recognizing the importance of a good quality sleep to alleviate stress while establishing a measurable relationship between stress and sleeping habits. A system that analyzes the sleeping habits by continuously monitoring the physiological changes that occur during rapid eye movement (REM) and non-rapid eye movement (NREM) stages of sleep is proposed in the current work. In addition to the physiological parameter changes, factors such as sleep duration, snoring range, eye movement, and limb movements are also monitored. The SaYoPillow system is processed at the edge level with the storage being at the cloud. Not having to compromise the user's privacy, SaYoPillow proposes secure data transmission for both uploading and retrieving, and secure storage and communications as an attempt to reduce malicious attacks on healthcare. A user interface is provided for  the user to control data accessibility and visibility. SaYoPillow has 96\% accuracy which is close to other existing research works. However, SaYoPillow is the only work with security features as well as only work that considers sleeping habits for stress. 
\end{abstract}



\section{Introduction}
\label{Sec:Introduction}

Stress can be defined as a state of mental or emotional strain due to unavoidable or demanding circumstances. Stress can be also be defined as a specific strain on the human body caused by various stressors. The standard origin for human stress is abbreviated as N.U.T.S. which is \textit{Novelty} to a situation or experiencing something new, \textit{Unpredictability} to sudden unexpected changes, \textit{Threat} to the ego or the survival of the individual, and \textit{Sense} of control where there is little or no control over the situation \cite{Lupien_LaRevue_2009}. 

Stress is generally categorized into distress or eustress, where distress is considered as negative stress and eustress is considered as positive stress. Stress is also categorized in terms or duration in three different categories: acute stress, episodic acute stress and chronic stress. Acute stress is mostly short and brief stress, while episodic acute stress is the repetition in the frequency of occurrence of acute stress. Chronic stress is categorized as follows: the wear and tear of stress response for being activated more than once for a small period of time, the habituation of stress to the human, failing to go back to the normal state from the stress state, prolonged exposure and inability to respond normally to stress, i. e. inadequate response. Prolonged exposure to chronic stress has many health hazards which include obesity, insomnia, diabetes, depression and sometimes even cancer \cite{Khan_JTEHM_2019}.

Stressors are anything that causes the human body to release stress hormones. Stressors are categorized as physiological, psychological, absolute and relative. Physiological or physical stressors are those which put strain on the human body like environmental temperature, injuries, chronic illness or pain. Psychological stressors are the events, situations, individuals or anything which are interpreted as negative or threatening. Absolute and relative stressors are defined based on the exposure of people to the intensity of the situation i.e., absolute stressors are those that everyone is exposed to, for example natural calamities, 9/11 or COVID-19, while relative stressors are those that only some people are exposed to like work pressure, exam pressure, traffic, and insomnia \cite{Schneiderman_Annu_2005}.

\begin{figure}[htbp]
	\centering
	\includegraphics[width=0.80\textwidth]{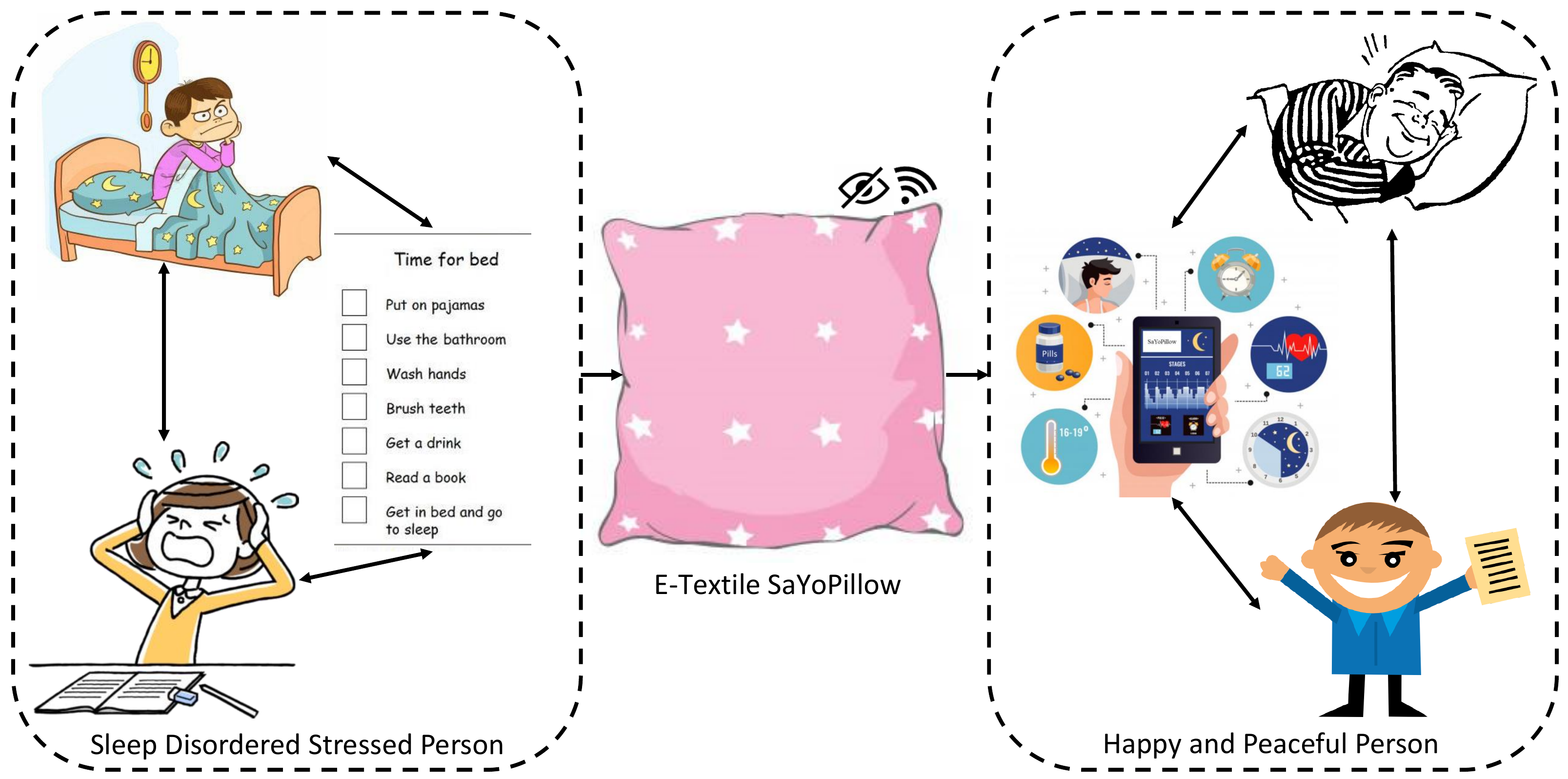}
	\caption{Proposed SaYoPillow as a Consumer Electronics of e-Textile based Pillow.}
	\label{fig:Device_Prototype_of_SaYoPillow}
\end{figure}

When exposed to stress due to various stressors for longer periods of time, humans develop lack of adaptation which can have major impacts on relationships, work, health and on the self by producing emotional breakdowns. Having an ability to build a self-monitoring system to tackle these stressors is very important. Incorporating these methods in the Internet of Medical Things (IoMT) creates a path for the users to adapt. A way to control and monitor the physiological stress and stress through food habits using the IoMT is presented in \cite{Rachakonda_TCE_2019, Rachakonda_TCE_2020} respectively. An attempt to control and monitor stress variations due to lack of sleep is the focus this work. The better the quality of sleep, the lower the stress levels \cite{Han_EN_2012}.

Protecting electronic health data is an important element in smart healthcare as it requires collection, storage and use of large amounts of sensitive personal information. The Health Insurance Portability and Accountability Act proposes not only safeguarding the patients' information, providing integrity, confidentiality, individuality, respect, dignity and worth but also maintaining a risk management algorithm to address in case of vulnerability. However, the number of breaches in the Healthcare Services are increasing at a high rate. There were 15 million and 32 million patient records which were compromised with 503 breaches in 2018 and 2019, respectively \cite{Koczkodaj_Iranian_2019}. Thus an attempt to decrease the rate of increase in lack of security has also been addressed in this work.

\begin{figure}[htbp]
	\centering
	\includegraphics[width=0.80\textwidth]{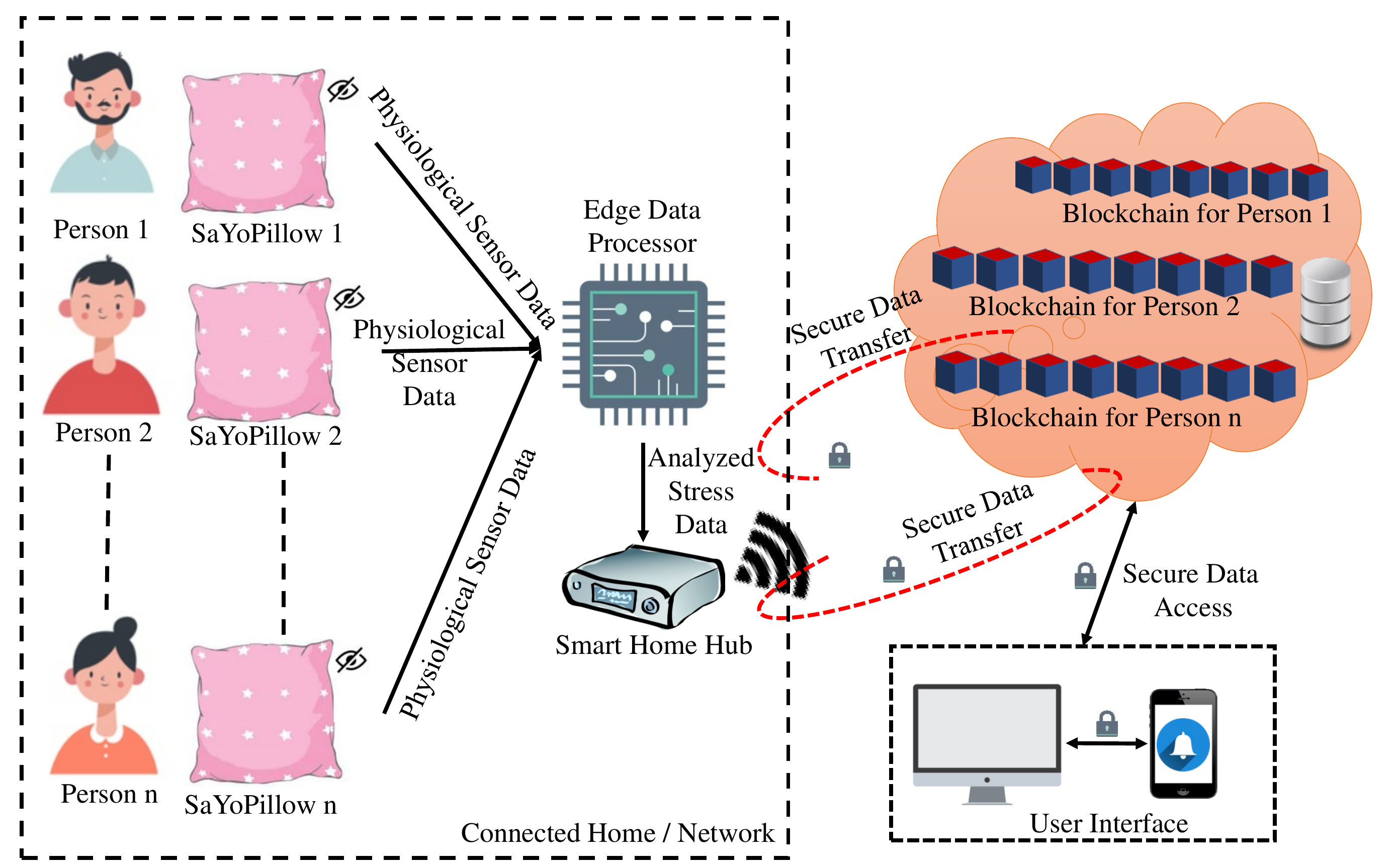}
	\caption{Security Privacy Aware SaYoPillow in Internet-of-Medical-Things (IoMT) based Healthcare Cyber-Physical System (H-CPS) for Smart Healthcare.}
	\label{fig:Schematic_Representation_of_SaYoPillow}
\end{figure}

Thus, we propose SaYoPillow, where we monitor and control the stress levels of a person during the sleep period. By securely monitoring the data during sleep helps not only the user to be prepared for the following day, but also helps in having an efficient sleep. A schematic representation of SaYoPillow is shown in Fig.  \ref{fig:Schematic_Representation_of_SaYoPillow}.

The rest of the paper is organized as follows: 
Section \ref{SEC:Research_Objective} presents the research questions addressed along with the motivation and objective of SaYoPillow. 
Section \ref{SEC:Related_Research} surveys the state of the art and lists the issues with existing research. Section \ref{SEC:Novel_Contributions} explains the vision of SaYoPillow, proposed solutions and novelties. Section \ref{SEC:SaYoPillow_Perspective} explains the relationship between stress and sleep along with its analytics. Section \ref{SEC:SaYoPillow_Models} describes the training methodology used in SaYoPillow. Section \ref{SEC:Stress_State_Analysis} describes the stress state detection during sleep and stress state prediction for the next day. Section \ref{SEC:SaYoPillow_Blockchain} justifies the usage of the blockchain in healthcare. Section \ref{SEC:SaYoPillow_Implementation} presents the experimental implementation and  validation of SaYoPillow. Section \ref{SEC:Conclusion} concludes the paper and presents possible future research.

\section{Research Questions and Objectives addressed in SaYoPillow}
\label{SEC:Research_Objective}

\subsection{Research Questions}

The motivation to propose a system which performs Stress Detection, Prediction and Control using the sleeping behaviors of a person through SaYoPillow is to provide a potential solution for the driving questions which are: 

\begin{itemize}
	\item Is sleep an important factor for non-negligible stress?
	\item Can there be a non-invasive, low complex system to monitor the changes that occur while sleeping?
	\item Can this solution be optimized?
	\item Can this solution be a part of the network or a smart home?
	\item Will the data that is taken from the user be private? 
	\item Can this solution work on edge devices?
\end{itemize}

\subsection{Proposed Solution}

With the positive potential of developing a solution, SaYoPillow proposes a non-invasive, optimized, IoMT enabled system which detects the stress level variations during sleep based on the sleeping parameters, predicts the stress level behavior for the following day, provides automated control mechanisms, analyses the data at the user end (at IoT-Edge) and stores the data at the cloud end (at IoT-Cloud) with no compromise in the user's personal information integrity.

\subsection{Research Objectives}

The idea behind proposing SayoPillow is motivated by taking typical human behavior into consideration. The objectives behind our idea are explained below. 
\begin{enumerate}
		\item 
		User's Convenience. \\
		There are few wearables and applications that log the sleep data of the user, as mentioned in Section \ref{SEC:Related_Research}. However, there are none that have the user's convenience as the main motivation.  
		\item 
		No User Input. \\ 
		The state-of-art in this area addresses semi-automated methods when it comes to notifying the user, as explained in Section \ref{SEC:Related_Research} but the analysis of the factors considered in studying the sleep parameters should always be done by the user. Proposing a no user input, fully automated and response control system was our second objective.
		\item 
		User Education \\ 
		Educating the user on how important is the quality of sleep is considered in our system. Helping the user understand sudden physiological changes which may or may not have been given importance previously, to eliminate the chances of chronic stress and to establish the relationship with sleep quality and the moods of the person is our third objective. 
		\item 
		Comprehend the Phrase ``Smart-Sleeping.'' \\
		``Smart-Sleeping'' can be defined as sleep that is complete and which meets the ideal body requirements during sleep. Knowing the difference between just sleeping and making the most of the sleep is what makes it smart. 
			
\end{enumerate}

\section{Related Prior Works and Research Gap}
\label{SEC:Related_Research}

\subsection{Related Prior Research}

The consumer electronics literature provides advancements in technologies that could be useful for healthcare applications \cite{Adedoyin_TCE_2007, Kim_TCE_2010, Kawase_TCE_2020}. Video advancements and 3D effects in images are proposed in \cite{Adedoyin_TCE_2007} which can be used in any image related application. \cite{Kim_TCE_2010} also helps in improving the image quality and decrease the error rate by increasing the pixel quantity. Enhancements in automated speech recognition are another great addition as it is an important application with a potential to be used in further healthcare applications \cite{Kawase_TCE_2020}. \cite{Nath_CEM_2020} presents a review of existing stress detection research. A study shows that the muscle to muscle communication is effected whenever a person experiences stress \cite{Lin_ICCEC_2016}. Human stress while driving is addressed in \cite{Magana_ICCEB_2015}. A stress monitoring method using the data that is acquired by a smart watch is proposed in \cite{Ciabattoni_ICCE_2017}. Virtual Reality technology to monitor stress is proposed in \cite{Wiederhold_CEM_2018} for better health. A non-invasive stress monitoring method is proposed in \cite{Lawanot_ICCE_2019}. A stress monitoring system in students during classes is proposed in \cite{Sunthad_ICCE_2019} for better classroom experience. A stress monitoring system using cortisol as a bio-marker is proposed in \cite{Nath_ICCE_2020}. Another research to monitor stress by considering mouse and keyboard usage is proposed in \cite{Ciabattoni_ICCE_2020}. However, none of the articles mentioned above provide the concept of "Smart-Sleep", provide stress control mechanisms or provide secure data transfer and storage mechanisms. 

A study that establishes the relationship between emotions and stress is proposed in \cite{Gaffary_TAC_2020}. However, this study lacks the consideration of more possible features and also the importance of having a good quality sleep. A semi-automated system is proposed where stress is detected from studying brainwaves of a person is proposed in \cite{Mitrpanont_ICT_2017}. However, consideration of other physiological features is neglected. Also, during chronic stress or long term stress people adapt to the stress with no effect on the system. Another research which remotely monitors the respiratory rate of a person to detect stress is proposed in \cite{Shan_ACII_2018}. However just monitoring the respiration rate is not sufficient for accurate stress detection. Low frequency dielectric properties in human blood are studied to enhance the stress measurement efficiency in \cite{Abdalla_TNB_2011}. Similarly, the quality of EEG signal is enhanced to detect stress in \cite{Hu_TNB_2015}. However, these works do not explain the relationship between stress and sleep. A stress classification is proposed in \cite{Arsalan_JBHI_2019} using Muse headband \cite{Muse_2016} and a wearable is proposed in \cite{Gurel_SJ_2019} for the same. They don't focus much on stress detection and also lack the relationship between stress and good quality sleep. Using saliva, a study has been proposed to monitor cortisol levels \cite{Wu_JBHI_2019}. This work did not study the importance of sleep with the variations of cortisol levels. All these research works do not focus on secure transmission and storage of data, thereby ignoring the privacy issues.

\subsection{Related Consumer Products}

In \cite{Lee_IERPH_2018}, a study of sleep patterns involving participants and already existing wearables is given. A mobile application has been developed in \cite{Zhenyu_ICPCTH_2013} without using wearables for monitoring. There are wearables like the Xiaomi Mi \cite{xiaomi_band3}, Eversleep \cite{eversleep}, Dreem \cite{Dreem}, Fitbit \cite{fitbit_2016} and Muse \cite{Muse_2016} which help in monitoring the sleeping habits of a person. Xiaomi helps in assisting users to have a peaceful sleep by monitoring their breathing patterns. Eversleep tracks sleep patterns using the pulse of a person. Dreem and Muse do not consider various physiological parameters that vary during sleep but they stimulate the slow wave signals to regulate sleep. Fitbit monitors the heart rate of a person to assess the quality of sleep. Some of the non-wearables that are available for regulating sleep are Beddit \cite{Beddit}, SleepScore Max \cite{watson_2018}, and Eight \cite{eight}. Beddit and SleepScore Max focus on the snoring rate of a person and Eight tracks parameters such as temperature, breathing rate, humidity, heart rate and light levels for better sleep. However, most of the wearables and non-wearables which are mentioned here do not monitor all the physiological changes during sleep and they do not provide a relationship with stress and sleep. A few of the most commonly used devices are given in Table \ref{Table: Sleep Trackers}. All these wearables and non-wearables do not have a secured method for the transmission and storage of data that is used for detection.

\begin{table*}[htbp]
		\caption{Sleep Trackers}
		\label{Table: Sleep Trackers}
		\centering
		\begin{tabular}{|p{2.8cm}p{2.2cm}p{4.5cm}p{4.5cm}|}
			\hline \hline
\textbf{Consumer Products} &	\textbf{Approach} & \textbf{Features} & \textbf{Drawbacks} \\
			\hline 
			\hline
			Fitbit \cite{fitbit_2016} &  Wearable &  Heart rate monitor, sleep stages monitor. Has techniques to improve the sleep score.  & Relationship between stress and sleep is not discussed.  \\
			\\
			SleepScore Max \cite{watson_2018} & Non-wearable & Invisible radio wave sleep tracking &  Does not manage stress with sleep. \\
			\\
			Nokia Sleep \cite{Nokia_2016} & Non-wearable & Uses Ballistocardiography sensor & Does not explain the relationship with stress with sleep. \\
			\\
			Xiaomi Mi Band 3 \cite{xiaomi_band3} & Wearable & Pulse Monitor & No information on importance of quality sleep. \\
			\\
			Eversleep \cite{eversleep} & wearable & Snoring and breathing interruptions & No explanation on the relationship between stress and sleep. \\
			\\
			Beddit \cite{Beddit} & Non-wearable & Monitors snoring & Doesn't consider other possible features. \\
			\\
			Eight \cite{eight} & Non-Wearable & Humidity, temperature, heartbeat, breathing rate & No data on how it is important to have a good sleep. \\
			\\
			Dreem \cite{Dreem} & Wearable & Simulates slow brain waves & It doesn't consider other features; Does not manage stress with sleep. \\
			\\
			Muse \cite{Muse_2016} & Wearable & Simulates brain waves & No understanding of the importance of quality sleep. \\
			\hline
			\hline
		\end{tabular}
\end{table*}

\subsection{Issues with the Existing Solutions}

It is evident that the state of art today does not address the relationship between stress variations during the day and sleeping behaviors at night. The major issues that are not properly addressed in most of the current research are:

\begin{itemize}
		\item Users are not educated on the importance of having a quality sleep and how it effects stress. 
		\item None of these solutions store personal data in a privacy assured manner.
		\item Automatic stress detection and classification methods were not provided. For example, most of the existing research works require participants for data acquisition and heavy machinery for the study to be completed. 
		\item Analysis of the variations of stress level with quality of sleep is not presented.  
		\item In most of the research, techniques to automatically control the variations of stress levels of the user are not provided. 
		\item A convenient stress detection wearable was not proposed. 
		\item Multiple features to detect stress are not considered.  
		\item Fully-automated systems were not proposed by any of the articles cited above.
\end{itemize}

\section{The Novel Contributions of the Current Paper}
\label{SEC:Novel_Contributions}

\subsection{Our Vision of SaYoPillow}
\label{Section: Our vision}
\textbf{S}m\textbf{a}rt-\textbf{Yo}ga Pillow (SaYoPillow) is envisioned as a device that may help in recognizing the importance of a good quality sleep to the stress variations while establishing secure storage of the user's data. The sleeping habits are considered as an important factor to determine the stress levels of a person as sleeping disorder is a one of the primary stressors, along with other factors, such as emotional imbalance and eating disorders \cite{Han_EN_2012},\cite{Kim_bsm_2007}. Studies show that trying to normalize sleep behavior by regulating sleep duration and adjusting schedules helps in improving sleep quality  \cite{Colten_book_2006}. Having an ability to control stress by understanding its causes is more effective for long term stability. As explained above, sleep by itself is studied as a process involving changes in physiological parameters \cite{De_Sleep_1993,Gut_PR_2016}. Our vision in proposing SaYoPillow is to introduce a device, which can monitor these changes in the physiological parameters and predict stress behavior for the following day along with detecting the stress variations during sleep.

The proposed device prototype along with a detailed explanation of our vision, which demonstrates how a sleep disordered stressed person, trying to maintain a log is failing versus a happy person who has his physiological parameters monitored in a privacy assured connected way is stress free, is shown in Fig. \ref{fig:Device_Prototype_of_SaYoPillow}. This also indicates that SaYoPillow can be potentially manufactured using smart fabrics or electronic textiles.

\subsection{Proposed Solution through SaYoPillow}

SaYoPillow proposes a non-wearable device which is capable of: 
{\begin{itemize}
		\item Automatic continuous monitoring of physiological parameter variations during sleep with no user input. 
		\item A method which educates the users to fully attain and understand the importance of good quality sleep. 
		\item Automatic stress control mechanisms during sleep. 
		\item Suggesting various stress control mechanisms for the following day stress level predictions. 
		\item An interface representing the stress state analysis of the user during sleep by presenting the physiological signal parameters considered in SaYoPillow.
		\item Processing the information or data on an edge device while secure data storage is done in the cloud.
		\item Providing the users with ways to understand their detected past and predicted future stress state variations. 
	\end{itemize}

\subsection{Novel Contributions of SaYoPillow}
The novel contributions of this paper are: 

\begin{itemize}
	\item A continuously monitoring device which gets activated only when a person is lying on the bed to provide better battery life.
	\item An automatic stress detection and prediction method with no manual input. 
	\item Taking the general nature of humans into consideration, proposing an automated approach which takes the time that is spent by the user on bed to drift into sleep. 
	\item A five level status on detected stress and predicted stress is proposed based on sleep data. 
	\item Providing a fully-automated edge level device with secure data transfer to the privacy assured IoT-Cloud storage.
	\item Transferring the privacy assured data securely from the IoT cloud to the user interface for  user feedback. 
\end{itemize}

By reducing sleeping disorders and by having privacy assured secured data transfer and storage, this study can be a major advancement in the field of smart health care. The detailed explanation of SaYoPillow is represented in Fig. \ref{fig:Detailed_Representation_of_SaYoPillow}. The transition of the user from tossing and turning, trying to sleep and slowly drifting into sleep is represented here. This data is processed and analyzed in the edge device and the analyzed data is securely sent to the IoT-Cloud for storage and as a platform to securely transmit the data to the user interface. 

\begin{figure*}[htbp]
	\centering
	\includegraphics[width=0.85\textwidth]{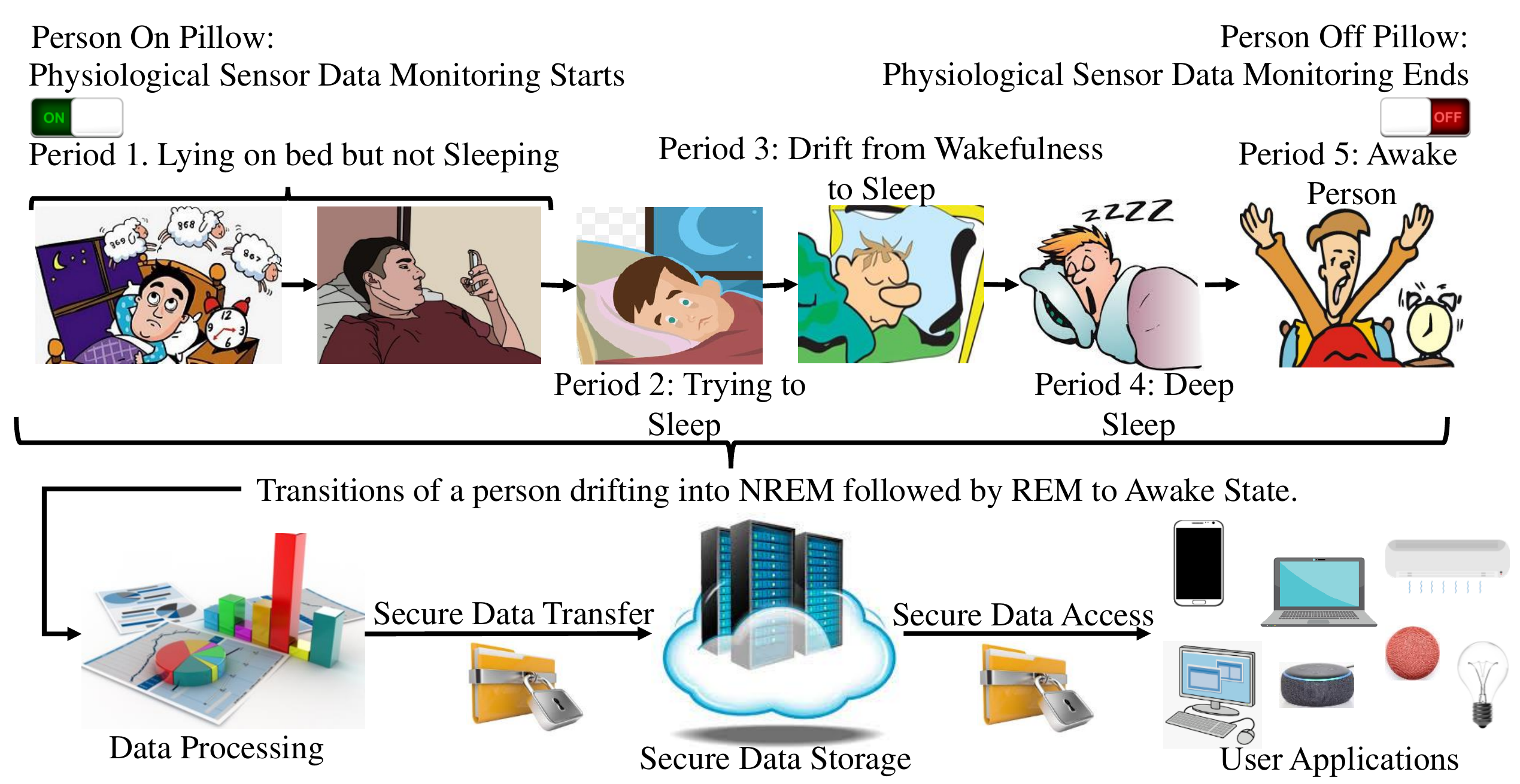}
	\caption{Detailed Representation of SaYoPillow.}
	\label{fig:Detailed_Representation_of_SaYoPillow}
\end{figure*}

\section{SaYoPillow: Proposed IoMT-Based Approach for Stress Detection and Prediction}
\label{SEC:SaYoPillow_Perspective}

The architecture of the SaYoPillow is represented in Fig. \ref{fig:Architectural_View_of_SaYoPillow}. Here, the privacy ensured secured pillow gets the user's data when the user puts pressure on the pillow. This will enable the monitoring of the physiological sensor signals considered. Once the data is collected, it is securely transferred to the edge device located remotely for the convenience of the user. The data is processed and the detection and prediction of stress state behavior of the user is analyzed. This analyzed information, with the use of the Internet is securely sent to the IoT-Cloud for storage. SaYoPillow is designed in such a way that for any third party application to retrieve the data, it could only be done securely maintaining the integrity and privacy of the user.

\begin{figure*}[htbp]
	\centering
	\includegraphics[width=0.95\textwidth]{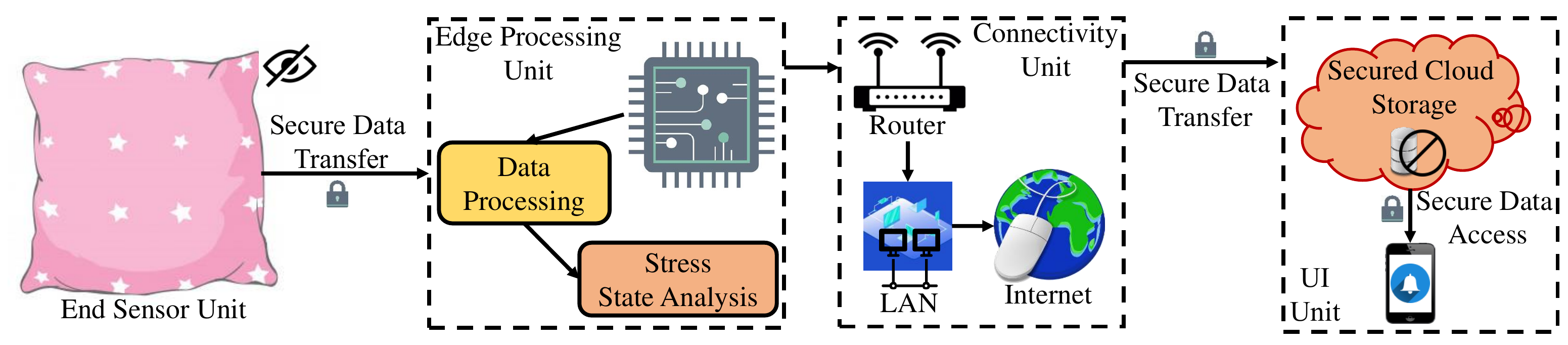}
	\caption{Architectural View of SaYoPillow.}
	\label{fig:Architectural_View_of_SaYoPillow}
\end{figure*}

\subsection{Relationship between Sleep and Stress}

\subsubsection{All about Sleep}
Sleep can be defined as a state in which the nervous system remains relatively inactive with eyes closed and relaxed muscles. Sleep is categorized to two stages: non-rapid eye movement (NREM) and rapid eye movement (REM) \cite{Robert_SleepMed_2007}. NREM occurs first and after a certain period it leads to deep sleep then followed by REM which is a relatively short period when compared to NREM. The sleep cycle of alternate NREM and REM stages takes an average of 90 minutes with a repetition of 4-6 times during the whole sleeping period. The NREM period is again classified in three stages where stage 3 is deep sleep stage. The pattern of the sleep cycle is represented in the Fig. \ref{fig:Sleep_Cycle}.  The REM sleep is observed when there is a shift in sleep stage to 1 or 2 in the NREM classification \cite{Silber_CSM_2007}. Soon after the REM stage, sleepers typically awaken from sleep. Unknowingly or without the sleeper's knowledge, a small amount of time during sleep is spent in a waking state. This waking state duration could be 0-1\% for females while 0-2\% for males \cite{Torbjorn_Sleep_2002}.   

\begin{figure}[htbp]
	\centering
	\includegraphics[width=0.65\textwidth]{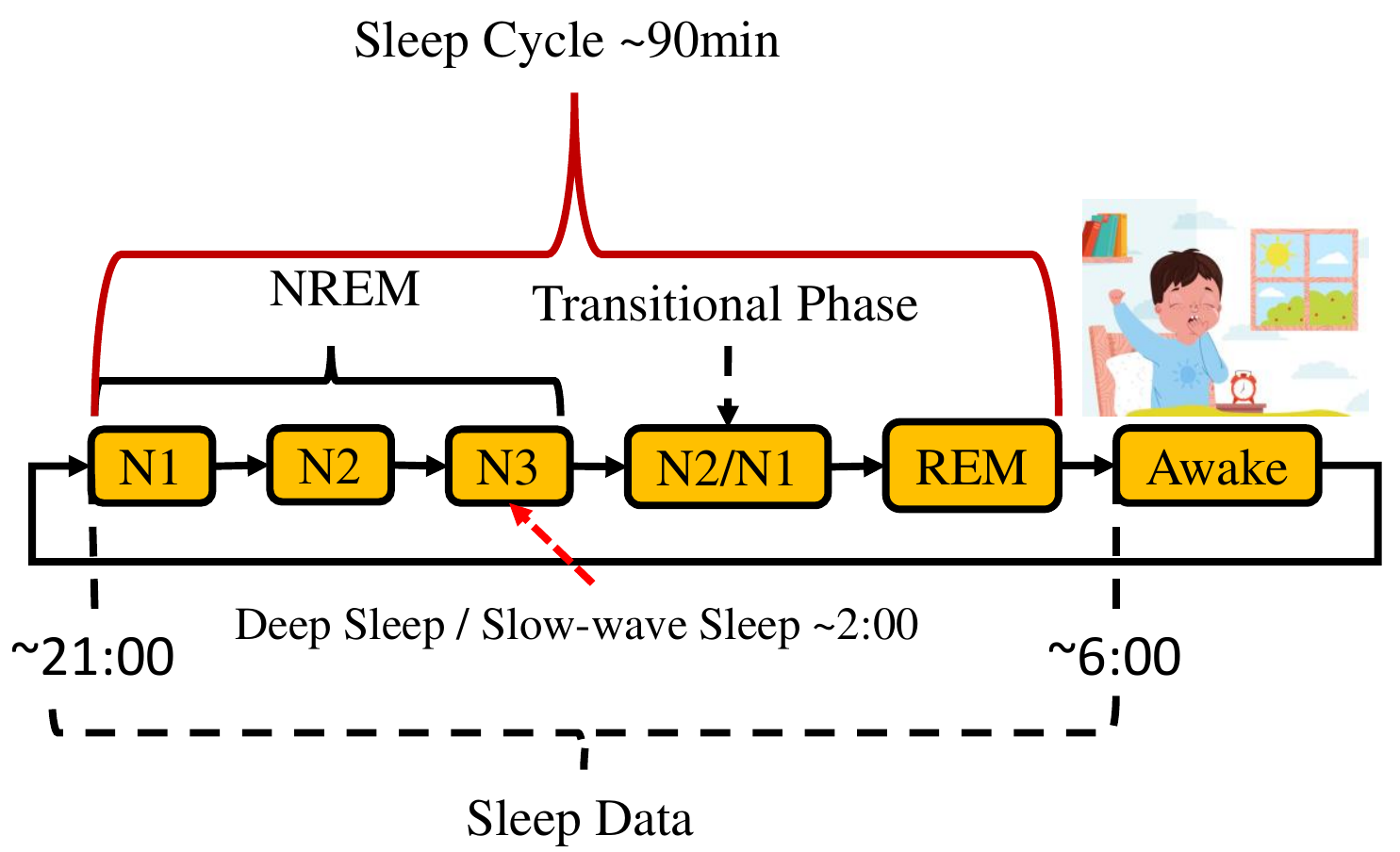}
	\caption{Detailed Explanation of Stages in Sleep Cycle.}
	\label{fig:Sleep_Cycle}
\end{figure}

\subsubsection{How to Study Sleep}

To identify sleep disorders, a sleep study called polysomnography is conducted. This helps in monitoring the sleep stages performing electroencephalography (EEG) to record brain waves, oxygen levels in the blood, heart rate and breathing as well as eye and leg movements by performing electrooculography and electromyography respectively. Studies based on polysomnography have observed that an anticipation of increased demand may disrupt some physiological indicators of sleep, which includes a decrease in the period of slow wave sleep and an increase in the heart rate \cite{Roebuck_PM_2013}. Mental capabilities such as learning, self-awareness and logical reasoning have been affected due to bad or low quality sleep \cite{Medic_NSS_2017}. There are studies that combine polysomnography with hormonal measurements to monitor the sleep patterns with physiological parameter variations \cite{van_P_2013}.

\subsubsection{Relationship between Stress and Sleep}
\label{Section:Stress and Sleep}

Studies show that the quality of sleep affects the quality of life during the day \cite{Kim_bsm_2007}. Disturbed sleep has been associated with increased instances of sickness, burnout syndrome, persistent psycho-physiological insomnia and higher risk of occupational accidents \cite{Akerstebt_Scand_2006}. Intrapersonal distress (experiencing anguish in one’s mind) and self-reported feelings of loneliness have been linked to lower sleeping efficiency and poor sleep quality \cite{Simon_NC_2018,Tavernier_DevP_2014}. Studies show that socializing and having people around results in having a good quality of sleep \cite{Colleen_ChInt_2006} while being lonely or not socializing reduces the efficacy of sleep \cite{John_PhySci_2002}. 

Frequent low quality or deprived sleep activates the stress hormone \emph{cortisol} in the human body \cite{Akerstebt_Scand_2006}. A recent study shows that lack of good quality sleep weakens the efficiency of the human body to produce enough proteins to fight infections and inflammations, thereby weakening the immune system \cite{Haspel_JCI_2020}. Post Traumatic Stress Disorder (PTSD) and systolic blood pressure are a few of the other negative effects on the human body due to low good quality sleep \cite{Franzen_PM_2012}. 
 



\subsection{IoMT-Cloud Secure Data Storage with Analytics at the Edge} 

The data which is collected at the end device is securely sent to the edge processor. The edge processor could be either located inside the pillow or could be a part of the network located remotely. The automatic data processing along with stress state analysis, as described in Sections \ref{data processing} and \ref{SEC:Stress_State_Analysis}, is done at the edge processor. This analyzed data is securely transferred to the IoT-Cloud for storage. Any third party applications that require the data, will have to securely retrieve it. 






\section{Proposed Training Methodology for Improving Smart-Sleeping Using SaYoPillow}
\label{SEC:SaYoPillow_Models}

\subsection{System Level Modeling of Physical Parameters with Stress}
\label{System level modeling}

The system level modeling of SaYoPillow is represented in Fig. \ref{fig:System_Level_Modeling_of_SaYoPillow}. Here, the data from the sleep study and the end sensors is taken for the analysis of stress. The data is sent to the edge platform which is connected using Wi-Fi. The data classification, extraction and processing is done along with the stress state behavior analysis. This analyzed data is sent to the IoT-Cloud for storage securely with the use of cryptographic techniques. The data retrieval is done from the cloud storage whenever needed with no compromise in the privacy of the users.  

\begin{figure}[htbp]
	\centering
	\includegraphics[width=0.85\textwidth]{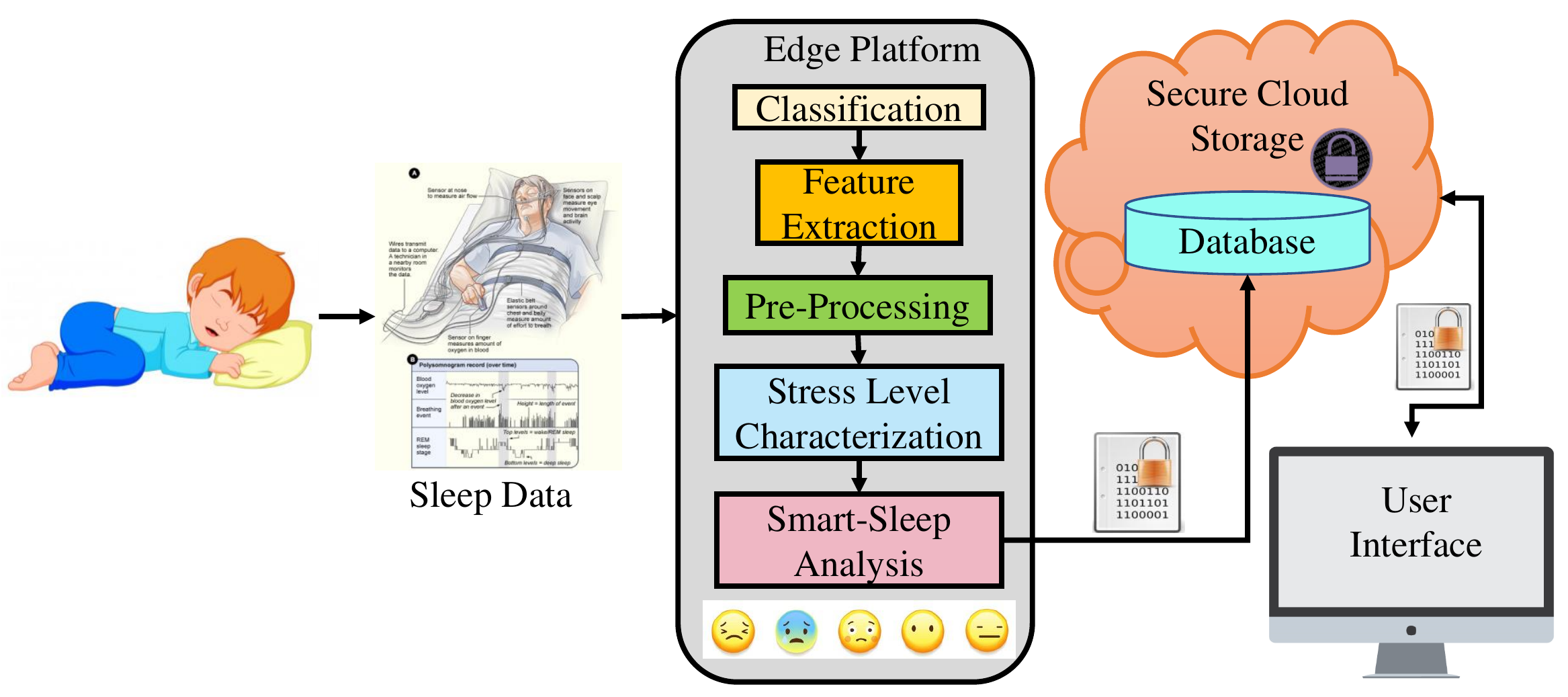}
	\caption{System Level Modeling of SaYoPillow.}
	\label{fig:System_Level_Modeling_of_SaYoPillow}
\end{figure}

\subsection{Modeling Physiological Data for Edge Platform}

SaYoPillow proposes the usage of physiological signals during sleep to access and understand stress behavior in humans. The type of physiological parameters considered, the methods to obtain these signals and the methods to process them is discussed in this section.

\subsubsection{Physiological Signal Classification and Extraction in SaYoPillow}
\label{Signal Literature}

Continuous monitoring of the physiological signals is done only when there is pressure applied on the pillow, thereby improving its battery life. Sleep can be considered as an active period which helps in developing optimal health and well-being by providing restoration and human body strengthening. In order to monitor the quality of sleep, SayoPillow proposes real time physiological signal monitoring by considering  parameters such as:
\begin{itemize}
	\item 
Number of hours of sleep,
	\item 
	Snoring range,
	\item 
	Respiratory rate range,
	\item 
	Heart rate range,
	\item 
	Oxygen in blood range,
	\item 
	Eye movement rate,
	\item 
	Limb movement rate, and
	\item 
 Change in body temperature.
\end{itemize}

Many physiological parameters tend to vary during the sleeping period. Snoring rate is measured in decibels as it is the sound that is produced when there is an unusual increase in the respiration rate per minute \cite{Levartovsky_J_2016}, \cite{De_Sleep_1993}. Snoring under 50dB can be considered normal, but when there is snoring at levels higher than 50dB, the chances of experiencing stress and other health issues are high \cite{Levartovsky_J_2016}, \cite{Hoffstein_Sleep_1996}. Similarly, the respiration rate, or the number of breaths taken per minute, is considered to be healthy if it is between 15-17 breaths per minute (bpm), and respiration rate greater than this is indicative of stress \cite{Gut_PR_2016}, \cite{Krieger_Sleep_1990}. In the same manner, the heart beats 5-10 times slower than the usual heart rate when a person is sleeping \cite{Cajo_PB_1994}. The heart rate is considered normal if the beats per minute (bpm) are 44-54 and considered abnormal if the bpm observed is greater than 54-64 \cite{Tochikubo_Hypertension_2001, Zema_P_1984}. Also, it is known that an adult should sleep for a minimum of 7 hours with at least 85\% of the time actually asleep; insufficient sleep will lead to stress \cite{Watson_Sleep_2015, Hirsh_SH_2015}.

During the transition to sleep, a person is drifting from being awake to falling asleep at the initial stage i.e., N1. Brain activity on the EEG is observed as low-voltage, mixed frequency waves. In N2, i.e., after 10 to 25 minutes in the initial cycle, the brain waves are observed to have low voltage, mixed frequency but higher speed. While in N3, the deep sleep stage, the brain waves are observed to have high voltage, and slow wave activity \cite{Colten_book_2006}. In a similar way, during sleep, the temperature of the human body tends to fall during the NREM stages. A slight raise in temperature resulted in a dramatic increase of good sleep quality according to \cite{Raymann_Brain_2008}. This fall in temperature is directly proportional to the period of duration in NREM stages. The lowest body temperature that a human can experience during sleep is 96.4 $^{\circ}$F a few hours before waking up. If there is any temperature that is less than 92.3 $^{\circ}$F and higher than 99.2 $^{\circ}$F during sleep, it indicates low quality sleep which may lead to stress \cite{Han_EN_2012}. Similarly, during the NREM period, the limbs of the human experience rapid movements or jerks which may affect the quality of sleep. These movements tend to occur every 20 to 40 seconds per one NREM episode. A situation at which there are at least four serial movements lasting 0.5-10 seconds for every 5-90 second duration can be termed as Periodic Limb Movements in Sleep (PLMS). Studies show that the Periodic Limb Movement Index (PLMI), which is the frequency of PLMS per hour of total sleep time, is 15 for 5-8\% of adults and increases with age \cite{Scofield_Sleep_2008}.  Any number of movements greater than fifteen can cause improper sleep and can lead to stress.

The stage in which the brain is most active is REM stage. There is a significant movement in the eyes, with an increase of heart rate, respiration rate and blood pressure. This is the stage when the body remains totally inactive while the brain remains active. Excessive amount of REM can lead to irritability, anxiety and even depression. It is advised to have 20-25\% of the total sleep duration in REM stage, which is approximately 90 minutes for 7-8 hour sleep \cite{Peever_CB_2016}. Healthy adults have oxygen levels in the blood around 93-98\% considering 95\% as normal . Oxygen levels are considered abnormal and can cause improper low quality sleeps leading to stress when they fall below 90\% \cite{Strohl_Webpage_2015}. 


\subsection{Proposed Fuzzy Logic Based Modeling}

Mamdani type fuzzy logic has been used to test the model of SaYoPillow. Eight physiological parameters are considered here with specific ranges listed in Table \ref{Table: Parameterized Ranging}. As there are 8 parameters and 5 sets of states, the total rules which can be generated are 792. This is because of the total number of combinations $(C)$ possible with 8 parameters $(p)$ with 5 input possibilities for each $(i)$ with repetitions among the $(i)$ are defined as the following expression:
\begin{equation}
C(p,i) = {p+i-1 \choose i} =\left (\frac{(p+i-1)!}{i! \times (p-1)! } \right)
\label{EQN:C}
\end{equation} 

The trained fuzzy logic system with a display of 30 rules is represented in Fig. \ref{fig:Rules_of_the_Fuzzy}. The input terms, i.e. the physiological parameters which can be varied for the stress state detection are represented in yellow with the output state in blue. The output state is defined as a number ranging from 0 to 5. 

\begin{figure}[htbp]
	\centering
	\includegraphics[width=1.02\textwidth]{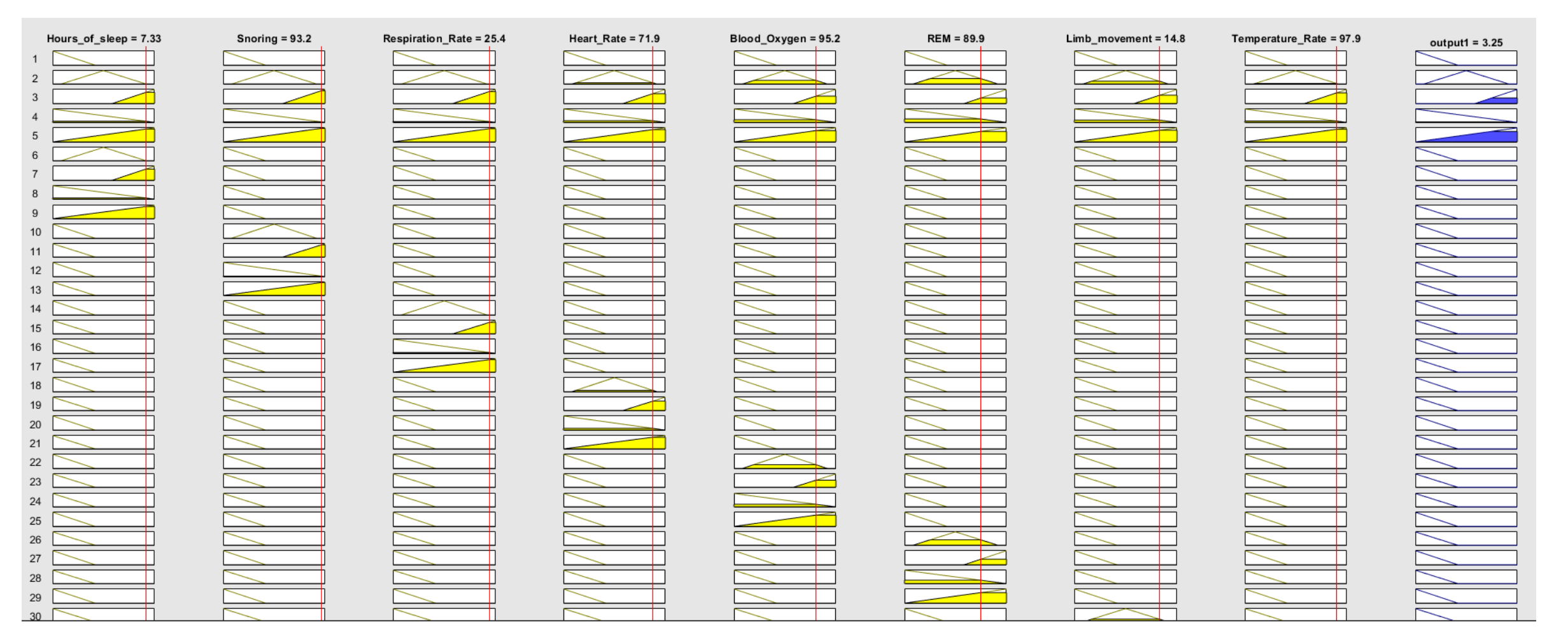}
	\caption{Rules of the Fuzzy System.}
	\label{fig:Rules_of_the_Fuzzy}
\end{figure}
 
The individual output range specification from 0 through 5 is divided among 5 stress state levels as shown in Table \ref{Table:Fuzzy Output Range Specification}.

\begin{table}[htbp]
	\caption{Fuzzy Output Range Specification}
	\label{Table:Fuzzy Output Range Specification}
	\centering
	\begin{tabular}{|p{5cm} p{2cm}|}
		\hline \hline
	\textbf{Stress State}  & \textbf{Output Range} \\
		\hline 
		\hline
		Low Stress State &  0.00-1.00    \\
		\\
		Medium Low Stress State & 1.01-2.00  \\
		\\
		Medium Stress State & 2.01-3.00  \\
		\\
		Medium High Stress State & 3.01-4.00  \\
		\\
		High Stress State & 4.01-5.00  \\
		\hline
		\hline
	\end{tabular}
\end{table}

\subsubsection{Data Processing in Fuzzy Logic Based Modeling}
\label{data processing}

Once the input data is received as the physiological parameters, it is compared to the range of parameters with which the stress state of the user can be detected from Table \ref{Table: Parameterized Ranging}. After the data is in the range, then the stress state is assigned according to the input parameters. The detailed flow of data processing with an example is provided in the Algorithm \ref{Algo1} while the stress state detection and prediction are detailed in Section \ref{SEC:Stress_State_Analysis}.

\begin{algorithm}[t]
	\begin{algorithmic}[1]
		\caption{Physiological Input Parameter Processing in Fuzzy Model used in SaYoPillow}
		\label{Algo1}
		\STATE Initialize the input random variables $HS$, $SR$, $RR$, $HR$, $BO$, $EM$, $LM$ and $TR$ to zero. 
		\STATE Initialize the output random variable $SLD$ to zero.
		\STATE Assign these variables to Number of hours of sleep, Snoring Range, Respiration Rate, Heart Rate, Blood Oxygen Levels, Eye Movement, Limb Movement and Temperature Rate to the initialized variables respectively.
		\STATE Compare the input data to the range of the parameters.
		\WHILE {the input data is in the range} 
		\STATE Assign value count of variables accordingly. For example, lets say the value of variable $HS$ is assigned to 6.
		\IF {variable is already assigned to a number}
		\STATE  Replace the existing variable with the latest input data i.e., if $HS$ is again detected as 7 rewrite the data from 6 to 7.
		\ELSE \STATE Continue to the next variable assignment i.e., assigning variable $SR$ to 90.
		\ENDIF
		\STATE Repeat the steps from 5 through 10 until all the variables are assigned. 
		\ENDWHILE
		\IF {any variable = 0}
		\STATE Repeat from step 4.
		\ENDIF
		\STATE Detect Stress State during sleep.
		\STATE Predict Stress behavior for following day.
	\end{algorithmic}
\end{algorithm}

A surface plot of the fuzzy design response is shown in Fig. \ref{fig:Surface_View_of_the_Fuzzy} which shows the variations of the output stress state with the variations in the input physiological parameters.

\begin{figure}[htbp]
	\centering
	\includegraphics[width=0.85\textwidth]{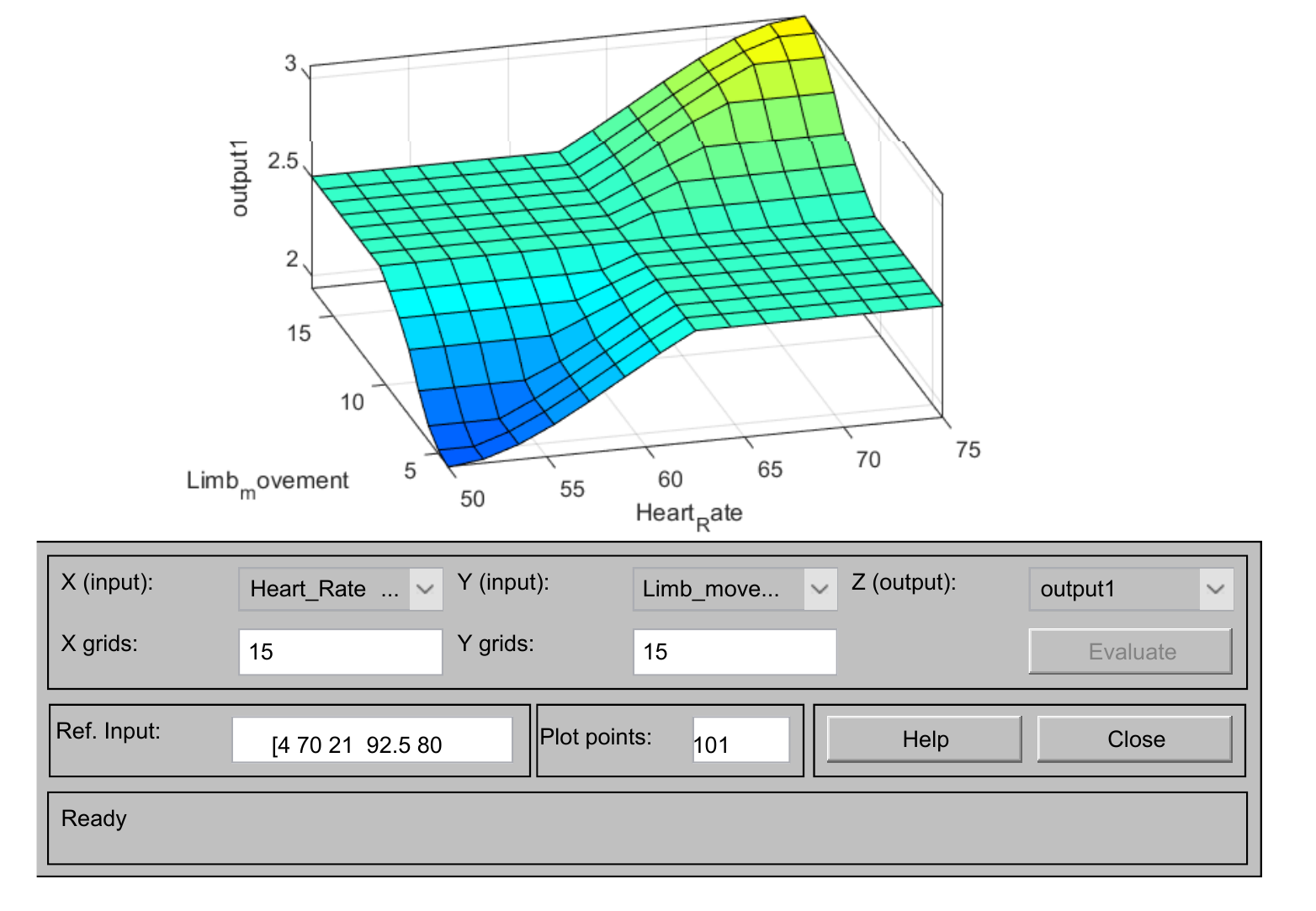}
	\caption{Surface Plot of the Fuzzy System Response.}
	\label{fig:Surface_View_of_the_Fuzzy}
\end{figure}

\subsection{Proposed Machine Learning Model}

In order to reduce the computation complexity and increase efficiency, a machine learning neural network model has been trained and tested in SaYoPillow. The dataset that is used for training and testing this model has been taken from the National Sleep Research Resource (NSRR). 

\subsubsection{NSRR Sleep Study Dataset}

For the process of training the ML model with baseline information, the dataset from NSRR is used. This study focused on heart and sleep disordered breathing. This study tests whether the sleep disorders are associated with heart diseases, strokes and hypertension. Thus this is considered for SaYoPillow as it proposes a sleep monitoring system to detect and predict stress. The study has the exam cycle taken from 6,441 individuals between 1995 and 1998. Out of these, raw polysomnography data is collected from 5,793 individuals. Polysomnograms were conducted at the houses of the participants with certified technicians' help \cite{Quan_Sleepdataset_1997,Zhang_JAMDATASET_2018}. As this study is basically conducted for heart and respiratory issues, not all information has been used in SaYoPillow. Out of many recordings, only the blood oxygen levels, respiration rate, heart rate, eye movement rate, body temperature, and limb movement were considered to analyze, detect and predict stress while the signal voltages of EEG waves are used for noticing the drift of sleep from awake to NREM. Additional features, such as snoring and number of hours of sleep is taken from the literature as explained in Section \ref{Signal Literature}.

\subsubsection{Data Processing in Machine Learning Model used in SaYoPillow}
\label{data processing}

15,000 samples were used in SaYoPillow out of which 13,000 samples are used for training while 2,000 samples are used for testing the model. In order to train and test with such a large dataset, TensorFlow is used. The dataset is fed into the model in CSV formatted file with 5 classes of stress level categories and approximately 3,000 samples per class. The batch size of the model to train the dataset is set as 32. Batched representation of features are represented in  Fig. \ref{fig:Batch_Representation_of_Data}. 

\begin{figure}[htbp] 
	\centering	
	\subfigure[Blood Oxygen Vs Snoring Range Plot]
	{\includegraphics[width=0.48\textwidth]{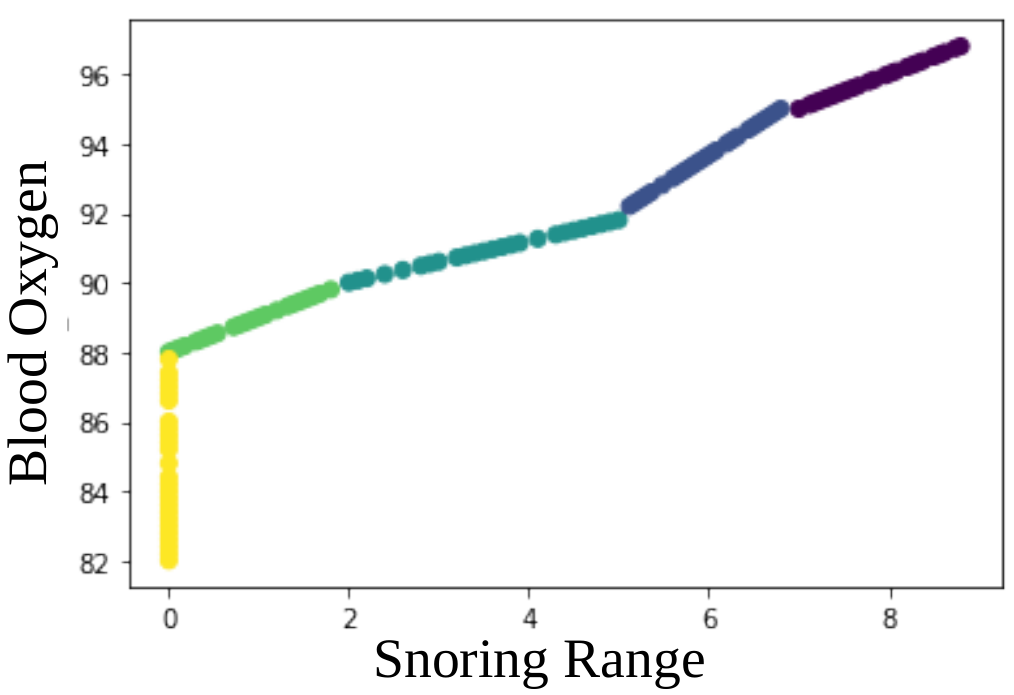}}
	\subfigure[Heart Vs Respiration Rate Plot]
	{\includegraphics[width=0.48\textwidth]{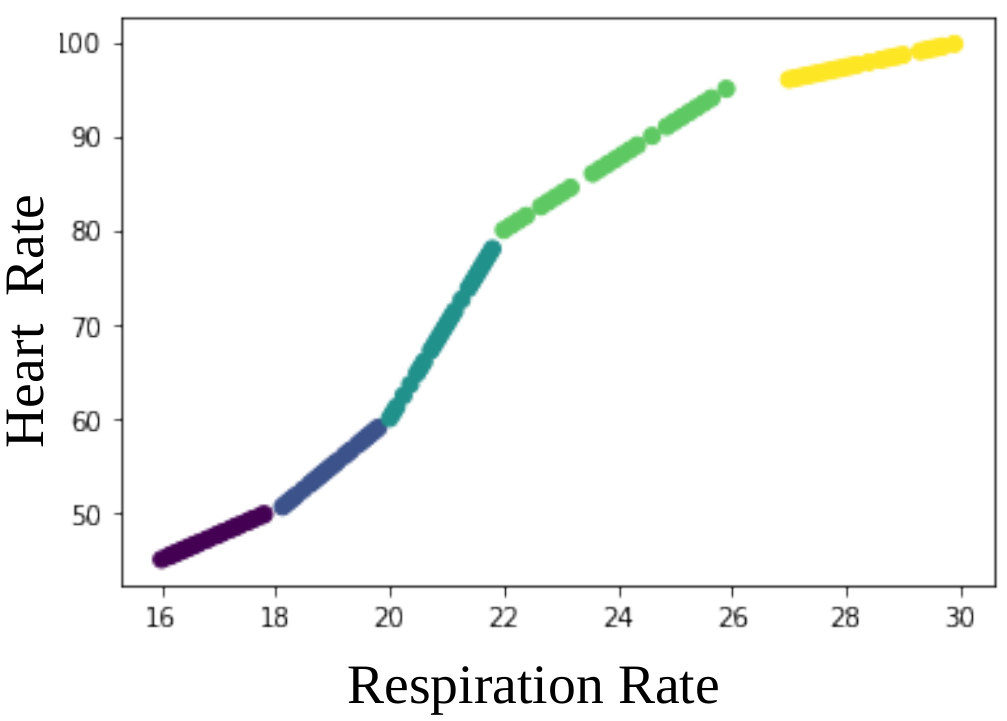}}
	\caption{Batch Representation of Data.}
	\label{fig:Batch_Representation_of_Data}
\end{figure}

A Fully-Connected Neural Network (FCNN) model is used to establish the relationship between the physiological parameters and stress levels as shown in Fig. \ref{fig:FCNN_Model_in_SaYoPillow}. An FCNN has neurons in one layer that receive input connections from every neuron in the previous layer. Here, we have 1 input layer, 2 hidden layers and 1 output layer. A sequential model with a linear stack of layers is created. The hidden layer has 10 neurons which are densely connected.

\begin{figure}[htbp]
	\centering
	\includegraphics[width=0.85\textwidth]{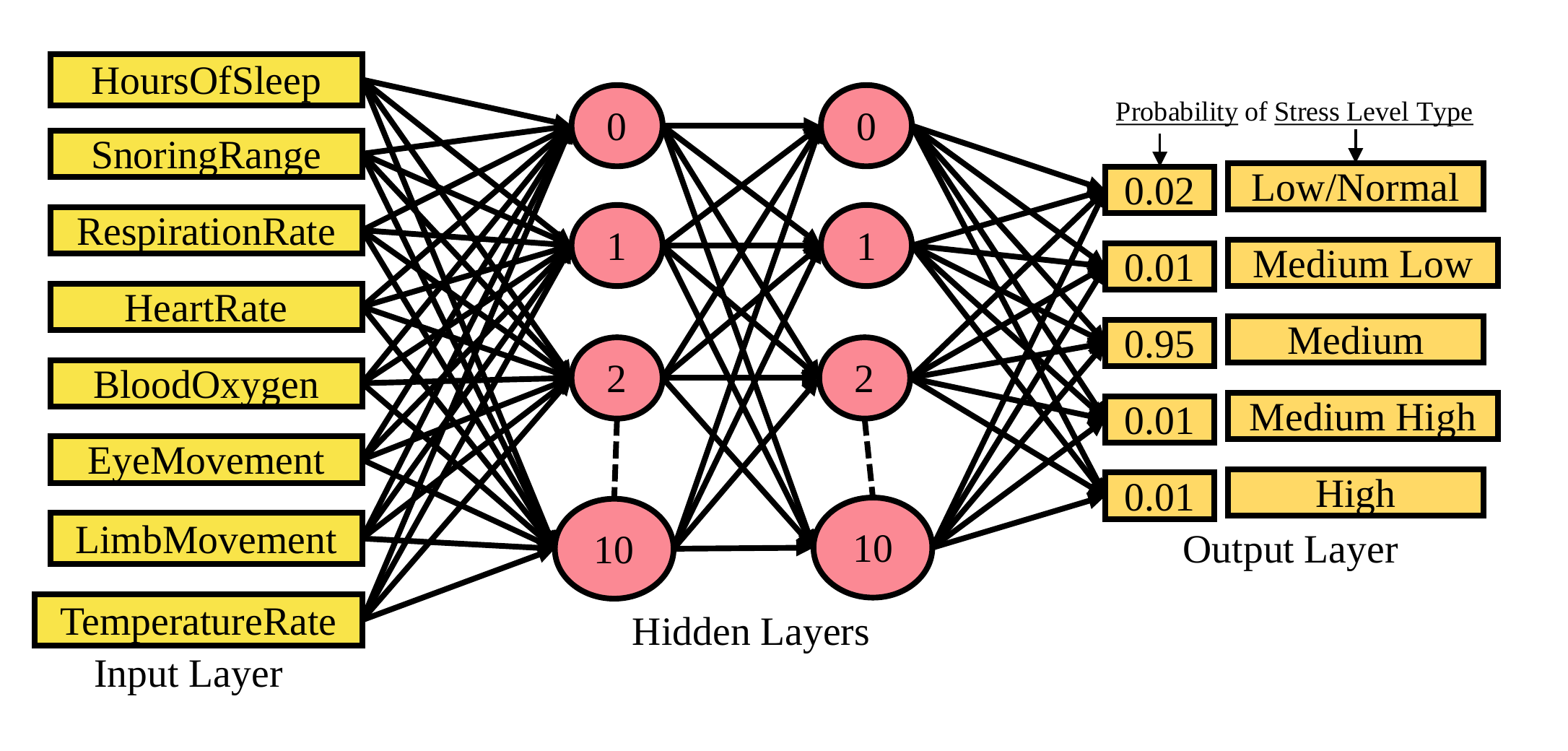}
	\caption{FCNN Model Representation in SaYoPillow.}
	\label{fig:FCNN_Model_in_SaYoPillow}
\end{figure}

After the data is fed to the model, it propagates through the hidden layers where the weighed inputs to that layer are calculated using Eqn. (\ref{EQN:zX}):
\begin{equation}
z(X)= \sum_{i=1}^{N} (\omega_i x_i + \omega_o),
\label{EQN:zX}
\end{equation}
where $X={x_1, x_2, \cdots, x_n}$ is the $n$-dimensional input, $z$ is the response of the neuron, $\omega_i$ are the weights for each input and $\omega_o$ is a constant bias.

This produces a net input, as given in Eqn. (\ref{EQN:nextlayer}) which is then applied to the activation functions to produce the output. The Rectified Linear Function is used as an activation function for the hidden layers represented in Eqn. (\ref{EQN:Relu}) while the Softmax function represented as Eqn. (\ref{EQN:softmax}) is used at the output layer. The predictions at the output layer are produced as given in Eqn. (\ref{EQN:nextlayer}). When the model is trained and fed with an unlabeled example, it yields 5 inferences at the output layer. 

\begin{equation}
f(x) = 
\begin{cases}
1 & x>1 \\
x & x=1 \quad \text{and} \quad 0 \\
0 & x<0 
\end{cases}
\label{EQN:Relu}
\end{equation}

\begin{equation}
p=\text{softmax}(\omega \cdot x+b),
\label{EQN:softmax}
\end{equation}
where $\omega$, $p$, and $b$ denote weight, predictor function and bias, respectively.

\begin{equation}
h_j= f( {(W)_j,i} \cdot {(x)_i} + {(b)_j,i} ),
\label{EQN:nextlayer}
\end{equation}
where $(W)_j,i$ is the weight matrix, $(b)_j,i$ the bias, and $f$ is the Rectified Linear Unit (ReLU) activation function.

For any neural network model, the loss, i.e, the factor indicating how bad the model is performing is very important. This is used to optimize the model. The  tf.keras.losses.SparseCategoricalCrossentropy function is used to calculate the average loss across the batch sizes of the dataset. Training the model is important as it will help in improving its inferences. The training steps defined in SaYoPillow are represented in Algorithm \ref{Algo2}.

\begin{algorithm}[htbp]
	\begin{algorithmic}[1]
		\caption{Training Steps for FCNN in SaYoPillow}
		\label{Algo2}
		\STATE Set the epoch value. Iterate each stress epoch. This epoch defines the number of times the dataset to loop. 
		\STATE Inside every repetition, iterate every example from training dataset by correlating its input features, i.e., the physiological parameters and the output labels, i.e., the stress levels.
		\STATE  Using these features and training datatset, make inferences.
		\STATE Compare the actual stress level outputs with the stress predictions from the previous step.
		\STATE Calculate loss at every epoch.
		\STATE Calculate the training data loss and accuracy in order to determine the overall efficiency. 	
		\STATE Update the variables to predict stress levels with the help of optimized algorithm using the \textit{Gradient Descent} algorithm.      
		\STATE \textit{Repeat the above steps for all the stress epoch count.}        
	\end{algorithmic}
\end{algorithm}

When this process of training is completed, the model interpretations are observed as follows: \\
Epoch 82000: Loss: 0.861, Accuracy: 92.500\% \\
Epoch 82050: Loss: 0.387, Accuracy: 96.333\% \\
Epoch 82100: Loss: 0.272, Accuracy: 96.667\% \\
Epoch 82150: Loss: 0.159, Accuracy: 97.167\% \\ 
Epoch 82200: Loss: 0.129, Accuracy: 96.333\% \\


In order to test the trained model, the test dataset is fed to the network. Once the model has been tested, the Testing Gradients with the specific output parameter are :\\
Example 0 prediction: High Stress (0.5\%)\\ 
Example 1 prediction: Low Stress (0.4\%) \\
Example 2 prediction: Medium Low Stress (1.8\%) \\
Example 3 prediction: Medium Stress (95.7\%) \\
Example 4 prediction: Medium High Stress (1.6\%) \\

Finally, the loss and accuracy of the model are observed to be approximately $<$1\% and $>$96\% respectively. These plots are shown in Fig. \ref{fig:Loss_and_Accuracy_of_SaYoPillow}.

\begin{figure}[htbp]
	\centering
	\includegraphics[width=0.80\textwidth]{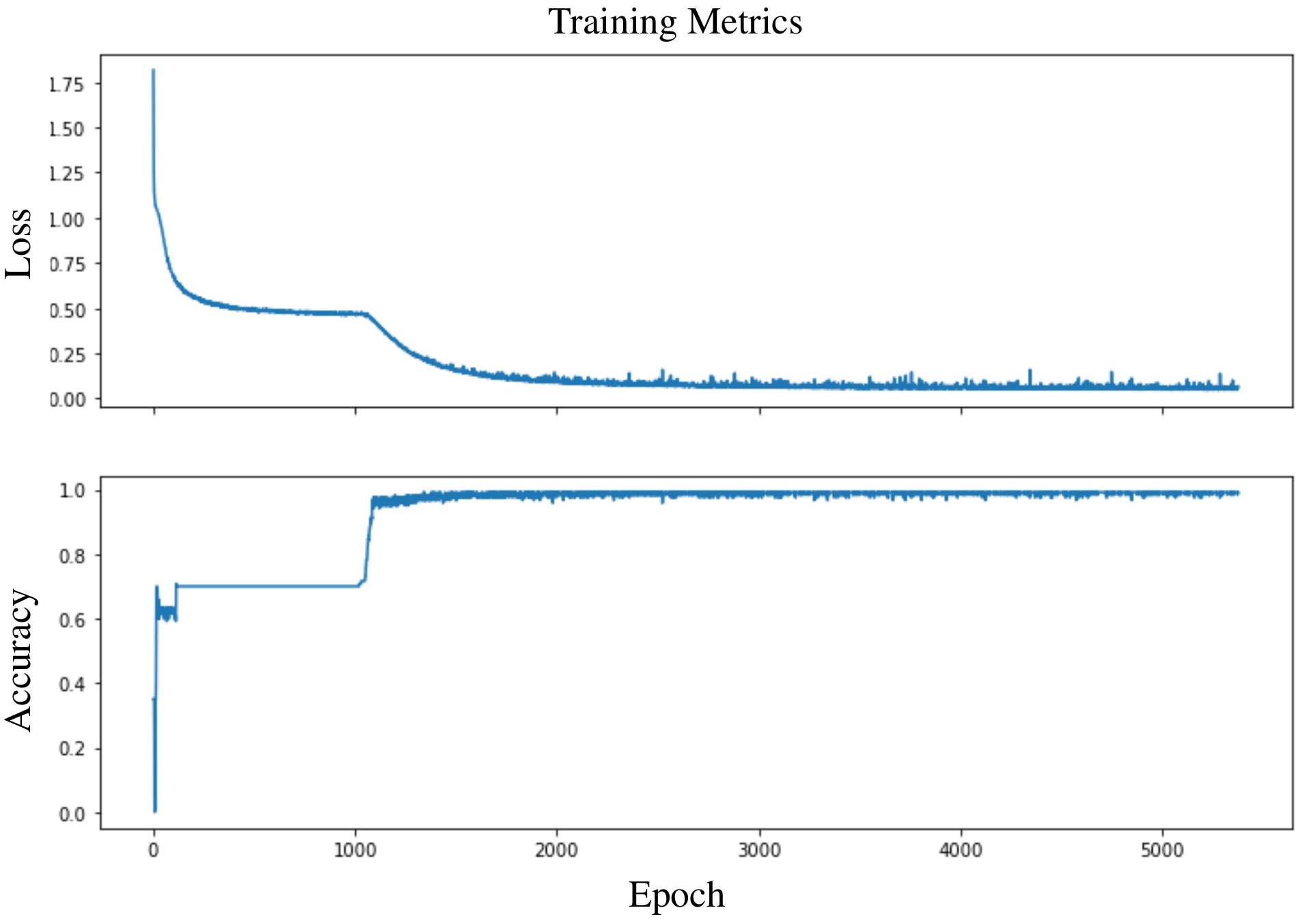}
	\caption{Loss and Accuracy of SaYoPillow.}
	\label{fig:Loss_and_Accuracy_of_SaYoPillow}
\end{figure}

\subsection{Metrics}

Metrics are used to monitor and measure the performance of th model. As in SaYoPillow we have used a model which is a classification related, the evaluation of performance is limited to few metrics which are discussed in this section. In the following discussion, $TP$, $TN$, $FP$, and $FN$ are defined as: 
\begin{itemize}
	\item True Positive ($TP$): A correct detection. This occurs when the detection using the Intersection Over Union (IOU) approach is greater than or equal to the set threshold value. 
	\item False Positive ($FP$): A wrong detection. This occurs when the IOU is less than the set threshold value. 
	\item False Negative ($FN$): This occurs when a ground truth is not detected. 
	\item True Negative ($TN$): This is the possible outcome where the model correctly predicts the ground false. This is not considered in object detection classifiers because there are many possible bounding boxes that should not be detected within an image.  
\end{itemize}

\subsubsection{Precision}

The ability of a model to classify only the relevant examples from the dataset is known as its precision ($P$).  
\begin{equation}
P = \left( \frac{TP}{TP+FP} \times 100\% \right).
\label{EQN:precision}
\end{equation}

\subsubsection{Recall}

The ability of a model to classify all the relevant examples from the predicted relevant examples is called recall ($R$): 
\begin{equation}
R = \left( \frac{TP}{TP+FN} \times 100\% \right).
\label{EQN:Recall}
\end{equation}

\subsubsection{Classification Accuracy}

The accuracy of a model can be defined as the ratio of correct predictions made by the model to all the predictions made by the model:
\begin{equation}
\alpha= \left (\frac{TP+TN}{TP+TN+FN+FP} \right) \times 100\%.
\label{EQN:A}
\end{equation} 

\subsubsection{F-1 Score}

F1 Score is a measure of test's accuracy and considers both precision and recall to make an appropriate choice. It is defined as harmonic mean of precision and recall and is represented as ($F1-score$):
   
\begin{equation}
F1-score= 2 \times { \left (\frac{P \times R}{P + R} \right) },
\label{EQN:F1}
\end{equation}

\section{Automatic Stress State Analysis and Control using Sleeping Habits in SaYoPillow}
\label{SEC:Stress_State_Analysis}

SaYoPillow proposes a system which can perform stress state detection during sleep and stress behavior prediction for the following day. The taken physiological signal data as input are processed and the stress state analysis is performed, which is later presented to the user. The stress state levels that are proposed in SaYoPillow, along with the detection, prediction and control methodologies to achieve ``Smart-Sleep'' are discussed in this Section. The flow of stress detection and prediction from processed data at edge is represented in Fig. \ref{fig:Proposed_Stress_Behavior_Analysis_in_SaYoPillow}. 

\begin{figure}[htbp]
	\centering
	\includegraphics[width=0.65\textwidth]{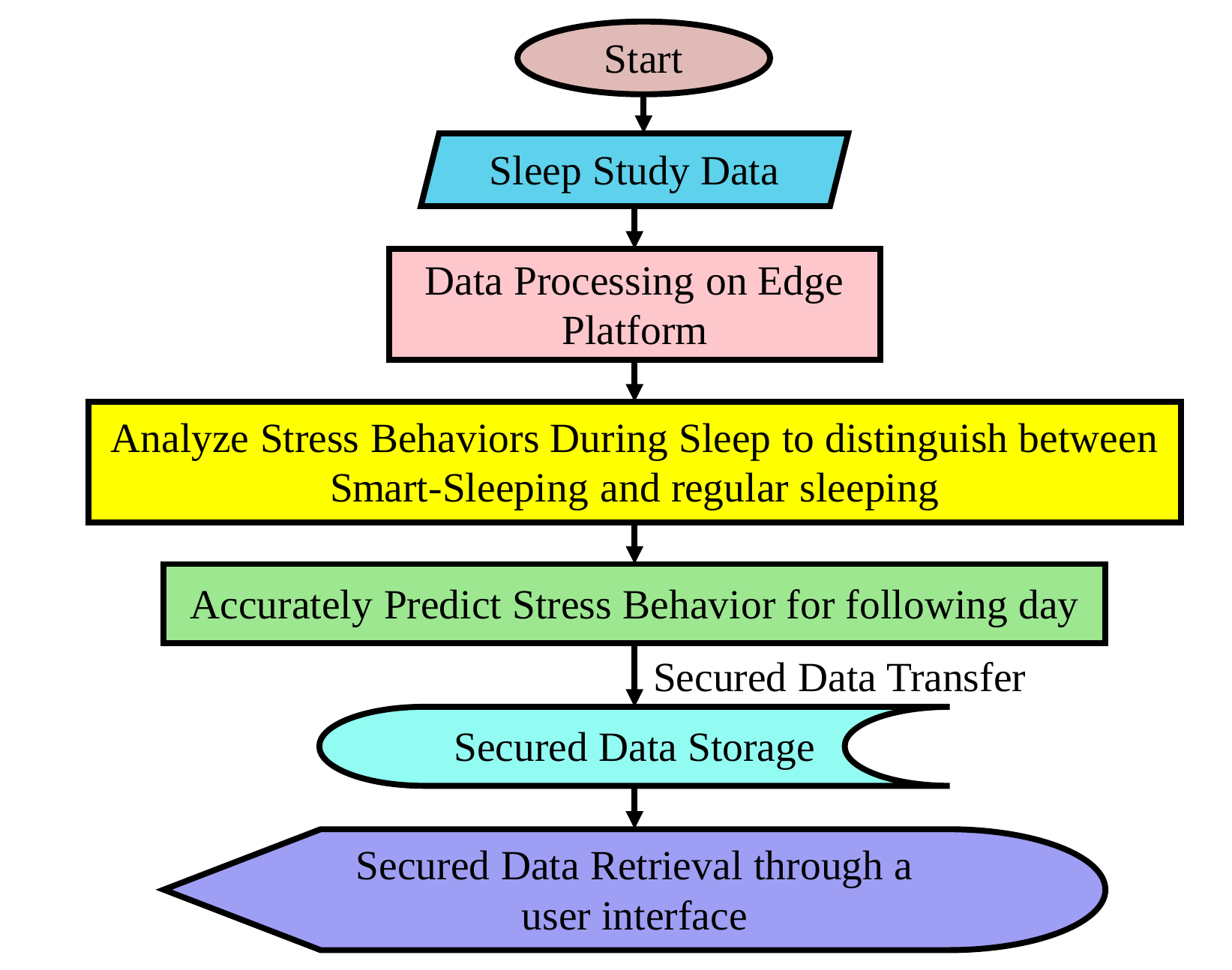}
	\caption{Proposed Stress Behavior Analysis in SaYoPillow.}
	\label{fig:Proposed_Stress_Behavior_Analysis_in_SaYoPillow}
\end{figure}

\subsubsection{Proposed Stress States for SaYoPillow}

After the data processing is done and the final stress state is detected from among the five proposed stress states of a person, it is displayed on the user interface available to the person. The available stress states are:
\begin{itemize}
	\item High Stress State
	\item Medium High Stress State
	\item Medium Stress State
	\item Medium Low Stress State
	\item Low Stress State
\end{itemize}

The low stress or normal stress is considered healthy while the high or medium high stress states are considered significantly unhealthy as explained in Section \ref{Signal Literature}. The characterization of each state with the physiological data is presented in Table \ref{Table: Parameterized Ranging}.

\begin{table*}[htbp]
	\caption{Stress Level Characterization with respect to Physiological Signal Data}
	\label{Table: Parameterized Ranging}
	\centering
	\begin{tabular}{|p{0.9cm}p{1.2cm}p{1.2cm}p{1.2cm}p{1.2cm}p{1.5cm}p{1.75cm}p{1.5cm}p{1.75cm}|}
		\hline \hline
\textbf{		Hours of Sleep} & \textbf{Snoring Range (dB)} & \textbf{Respiration Rate (bpm)} & \textbf{Heart Rate (bpm)} & \textbf{Blood Oxygen Range} & \textbf{Eye Movement Rate} & \textbf{Limb Movement Rate} & \textbf{Body Temperature ($^{\circ}$F)} & \textbf{Stress State} \\
		\hline 
		7-9 & 40-50 & 16-18 & 50-55 & 97-95 & 60-80 & 4-8 & 99-96 & Low/Normal  \\
		\\
		5-7 & 60-50 & 18-20 & 55-60 & 95-92 & 80-85 & 8-10 & 96-94 & Medium Low  \\
		\\
		5-2 & 60-80 & 20-22 & 60-65 & 92-90 & 85-95 & 10-12 & 94-92 & Medium  \\
		\\
		2-0 & 80-90 & 22-25 & 65-75 & 90-88 & 95-100 & 12-17 & 92-90 & Medium High \\
		\\
		$<$0 & $>$90 & $>$25 & $>$75 & $<$88 & $>$100 & $>$17 & $<$90 & High \\
		\hline
	\end{tabular}
\end{table*}

\subsubsection{Stress Behavior Analysis During Sleep for Smart-Sleeping}

The stress behavior of the person during sleep is considered as a real time data processing and analyzing methodology which is performed every 15 minutes. The process of this analysis is detailed in Algorithm \ref{Algo3}. 

\begin{algorithm}[htbp]
	\begin{algorithmic}[1]
		\caption{Stress Detection During Sleep in SaYoPillow}
		\label{Algo3}
		\STATE Declare and initialize timer $t1$, which is the time the user has applied pressure on pillow or is considered to lay on bed to 0.
		\STATE Declare and initialize timer $t2$, which is the time the user has slipped in to sleep to 0. 
		\STATE Declare and initialize timer $t3$, which is the time the user has woke up from sleep to 0.
		\STATE Declare string variables brain wave voltage $BWV$, frequency - cycles per second $CPS$ and Stress Level $SL$.
		\WHILE {$t1$ $\neq$ 0}
		\STATE Update $t1$ at instance 1. 
		\STATE Start monitoring and gathering physiological signal data which are $HS$, $SR$, $RR$, $HR$, $BO$, $EM$, $LM$ and $TR$. 
			\IF {$BWV$ == 'low' \&\& $t1$ == 30  \&\& 12$>$$CPS$$<$14}
			\STATE  Start $t2$ at instance 1; user drifting to sleep.
				\ELSIF {$t2$ $\neq$ to 0 \&\&  $BWV$ == 'high'  \&\& $CPS$$<$4 }
		 		\STATE Update $t2$ to current value at instance 2 ; user in deep sleep.
		 			\ELSIF {$t2$ $\neq$ to 0 \&\&  $BWV$ == 'high' \&\& $EM$ $\neq$ 0  }
		 			\STATE Update $t2$ at instance 3; user is in REM stage.
		 				\ELSE \STATE user is in light sleep. 
		 	\ENDIF
		 \IF {$t2$ $\neq$ to 0 \&\&  $BWV$ == 'low' \&\& $CPS$ $>$ 13 }
		 	\STATE Start $t3$ at instance 1; user woke up at alert state.
		 		\ELSE \STATE Start $t3$; user woke up relaxed.
		\ENDIF
		\IF {$t1$ == 0}
		\STATE Stop monitoring and gathering physiological signal data.
		\ELSE \STATE Repeat steps from 6 through 20.
		\ENDIF
		\ENDWHILE
		\STATE Compare the gathered data from every instance to the characterization Table.
		\STATE Detect the $SL$ of the user.
		\STATE Repeat the steps from 5 through 28  whenever a user is trying to sleep. 
	\end{algorithmic}
\end{algorithm}

\subsubsection{Stress Behavior Prediction for the Following Day in SaYoPillow}

The stress behavior prediction of the subject for the following day can be obtained by analyzing the stress levels observed during the night's sleep. The prediction analysis in SaYoPillow considers factors such as sleep latency in minutes, which is the period of time the user has taken to drift in to sleep, the quality of sleep and the total number of hours the user has actually slept. For a proper good quality sleep, the sleep latency should be in the range of 10 to 20 minutes \cite{Dement_book_1999}. Thus, as sleep latency increases, the quality of sleep decreases. 

The stress predictions are done by considering the time that the user has spent in different levels of stress during the actual sleep time. The longer the time the user experienced stressed sleep, the higher the probability of the user having a rough next day. The analysis is detailed in the Algorithm \ref{Algo4}.  

\begin{algorithm}[htbp]
	\begin{algorithmic}[1]
		\caption{Proposed Stress Prediction Analysis for the Next Day in SaYoPillow}
		\label{Algo4}
		\STATE Retrieve the updated variables $t1$, $t2$ and $t3$ at instances 1.
		\STATE Declare and initialize the variable sleep latency as $L$ to 0.
		\STATE Declare and initialize variable actual time slept to $ATS$ to 0.
		\STATE Declare and initialize string variable Stress Prediction to $SP$.
		\STATE Calculate the difference between the variables $t2$ and $t1$ to update $L$. 
		\STATE Calculate the difference between the variables $t3$ and $t2$ to update $ATS$.
		\WHILE {$SL$ is detected}  
		\IF {0.8*$ATS$ == 'M' $|$ $|$ 0.8*$ATS$ == 'H'$|$ $|$ 0.8*$ATS$ == 'MH' \&\& 35$<$ $L$ $>$45}
		\STATE $SP$ == 'MH'; user could experience rapid mood swings, irritability, sleeplessness and fatigue during the course of day.
		\ELSIF {0.6*$ATS$ == 'M' $|$ $|$ 0.6*$ATS$ == 'L'$|$ $|$ 0.6*$ATS$ == 'ML' \&\& 20$<$$L$$>$35 }
		\STATE $SP$ == 'ML'; user could experience some mood swings and tiredness during the course of day.
		\ELSIF { 0.8*$ATS$ == 'L' \&\& 10$<$$L$$>$20}
		\STATE $SP$ == 'L/N'; user could remain active and happy for the course of the day. 
		\ENDIF
		\ENDWHILE.
		\STATE Repeat the steps from 5 through 15 to predict stress every time.
	\end{algorithmic}
\end{algorithm}

After the detection and prediction, the next best thing to do after knowing the information is to have the capability of controlling it. 

\subsubsection{Proposed Stress Control Mechanisms in SaYoPillow}
\label{Sec. Stress Control}
The idea behind SaYoPillow introducing the phrase ``Smart-Sleeping'' is for the users to obtain the maximum benefit out of the natural process, sleeping. As discussed in Section \ref{Sec:Introduction} and in Section \ref{Section:Stress and Sleep}, the impact of stress on human body is non-negligible. In order to get control over the human body and face fewer health issues due to stress, we propose automated stress control when in stressful scenarios. As SaYoPillow has a connected approach to analyze and control stress with the use of the IoT, whenever there is a user on the pillow trying to sleep or $SL$ is detected high for 2 continuous instances, the edge processor which has the capability of connecting to devices can: 

\begin{itemize}
	\item Maintain the ambiance of the surroundings and environment by regulating room temperature with respect to stress level variation. 
	\item Control the lights when the user has spent 15 minutes in the sleep latency $L$ period.
	\item Play sleep music or peaceful tunes directly from the phone to sooth the thoughts of the user during sleep by connecting to a phone or any smart device.
\end{itemize}

Along with these automated control mechanisms, the user can be presented with many options including: to stay away from screens, take an evening bath, perform aroma therapy or light scented candles, read a book, and meditate \cite{Schneiderman_Annu_2005}. 

Depending upon the value of $SP$, the stress prediction factor, the user will be reminded through the user interface, to: 
\begin{itemize}
	\item Stay hydrated. 
	\item Eat good mood foods.
	\item Take walks periodically, and
	\item Display photos as notifications from the gallery.
\end{itemize}
 
After the stress prediction and detection are performed, the data along with physiological parameters gathered during those periods are safely and securely transferred to the cloud for storage. The transfer of the data is done by incorporating encryption and decryption techniques while the storage is provided in a fully secure, robust and access controlled blockchain. SaYoPillow proposes a storage system which not only encrypts the data while transferring but also provides security for information retrieval from any third party applications for user interfacing or for stress control mechanisms, thus maintaining the integrity of the user's vital data.

\section{Proposed Blockchain based Physiological Data Storage in SaYoPillow}
\label{SEC:SaYoPillow_Blockchain}

The selection of physiological parameters that are used in detecting and predicting stress in SaYoPillow is very significant and important. This vital information can be a part of the user's Electronic Health Records (EHR). This healthcare data has importance in not only analyzing the stress as in our application, but also can be used to monitor and diagnose different diseases in patients or during clinical trials. EHRs can be defined as the digitization of patient-level clinical information and are a part of smart healthcare systems. EHR systems help in exchanging patient information electronically between different entities involved in the healthcare system which provides efficient management of patient data and better health care. A typical healthcare system consists of many entities including doctors, patients, insurance providers, research groups etc., \cite{Beinke_JMIR_2019} and controlled access to the patient data is given as shown in Fig. \ref{fig:Overview_Application_with_stake_holders}. Using such EHRs will remove current cumbersome paper-based systems in healthcare, and this will result in better management for better healthcare \cite{Holbi_Symmetry_2018}. Some of the advantages of EHRs include:

\begin{itemize}
	\item Primary care details provided in real time using EHRs can help in preparedness of emergency rooms during health emergency.
	\item Repetition of tests can be avoided as previous test results and other medical information is present in the records.
	\item Patient follow-up and post-clinical care can be organized more efficiently.
\end{itemize}

\begin{figure}[htbp]
	\centering
	\includegraphics[width=0.75\textwidth]{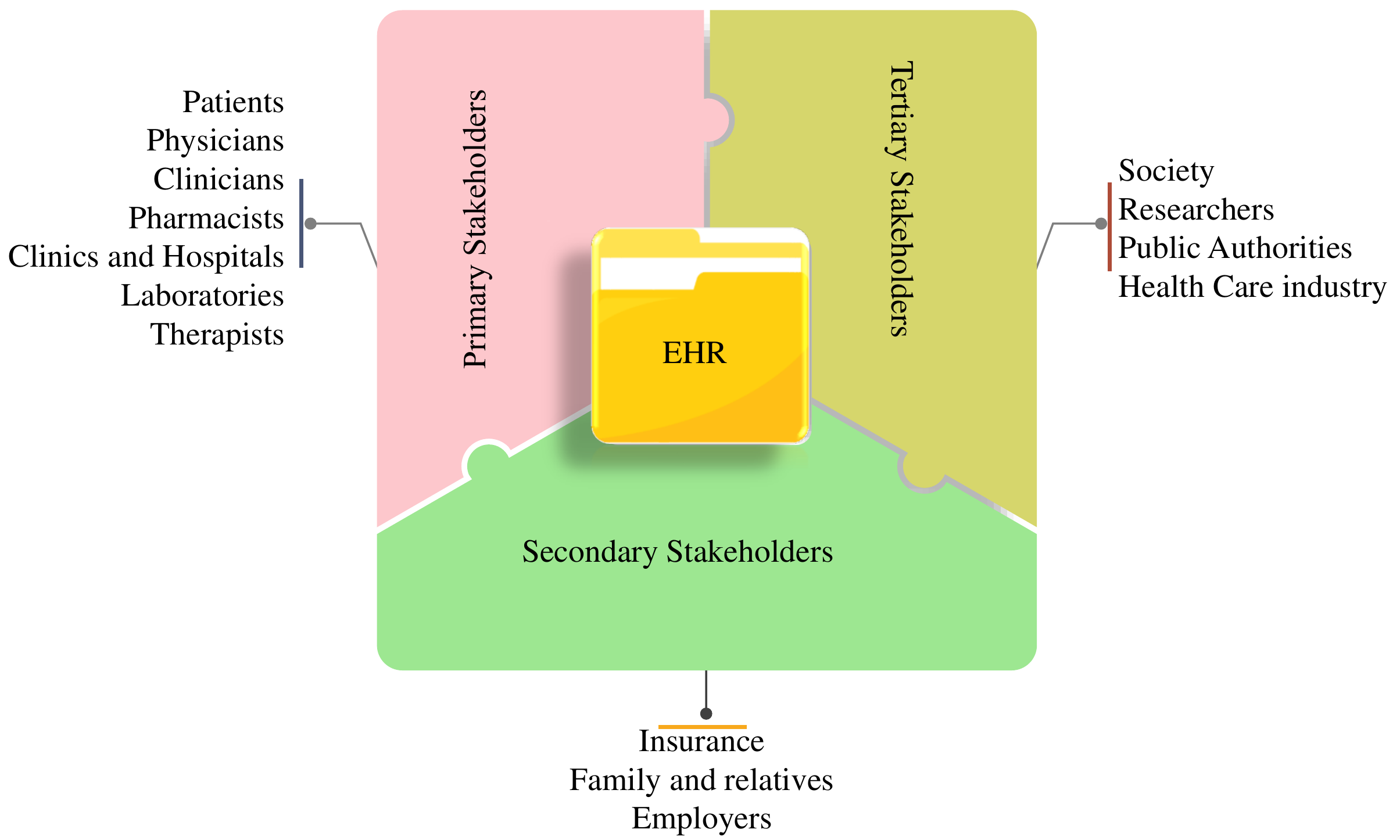}
	\caption{Overview Application with stake holders in SaYoPillow.}
	\label{fig:Overview_Application_with_stake_holders}
\end{figure}

Even though there are many advantages in replacing traditional healthcare systems with EHR systems, there are some limitations and challenges which still need to be addressed for implementing an efficient EHR system. Some of the major problems include information sharing, interoperability, data privacy, data security and availability. EHR systems as currently implemented are centralized and prone to single point of failure (SPOF). All these issues can be avoided by using a trusted decentralized mechanism such as the blockchain.

\subsection{All about the Blockchain}

The blockchain can be defined as a peer-to-peer communication network which provides a secure way of sharing data between untrusted entities working together in a system. The blockchain was first introduced in bitcoin as a financial solution for removing central authorities like banks and exchanges. This implementation has solved double spending problems and achieved anonymity in financial transactions \cite{Nakamoto_Cryp_2009}. It has shown very promising use cases in many different sectors. The IoT is one of such field which has  benefited from this new technology. The blockchain in its simplistic form can be defined as a ledger of transactions which keeps all the information in chronological order and secured.  Transactions are hashed and grouped together into blocks, all the blocks are connected chronologically using block hashes, hence forming a chain of blocks. It works in peer-to-peer communication protocols and all the peers in the network will have their own copy of the data. Immutability, irreversibility, decentralization, persistence and anonymity are some of the characteristics of the blockchain. These characteristics, as shown in Fig. \ref{fig:Characteristics_of_Blockchain}, have made the blockchain suitable for using in smart healthcare, because data privacy, data security and robust access control mechanisms are needed in such applications.  

\begin{figure}[t]
	\centering
	\includegraphics[width=0.90\textwidth]{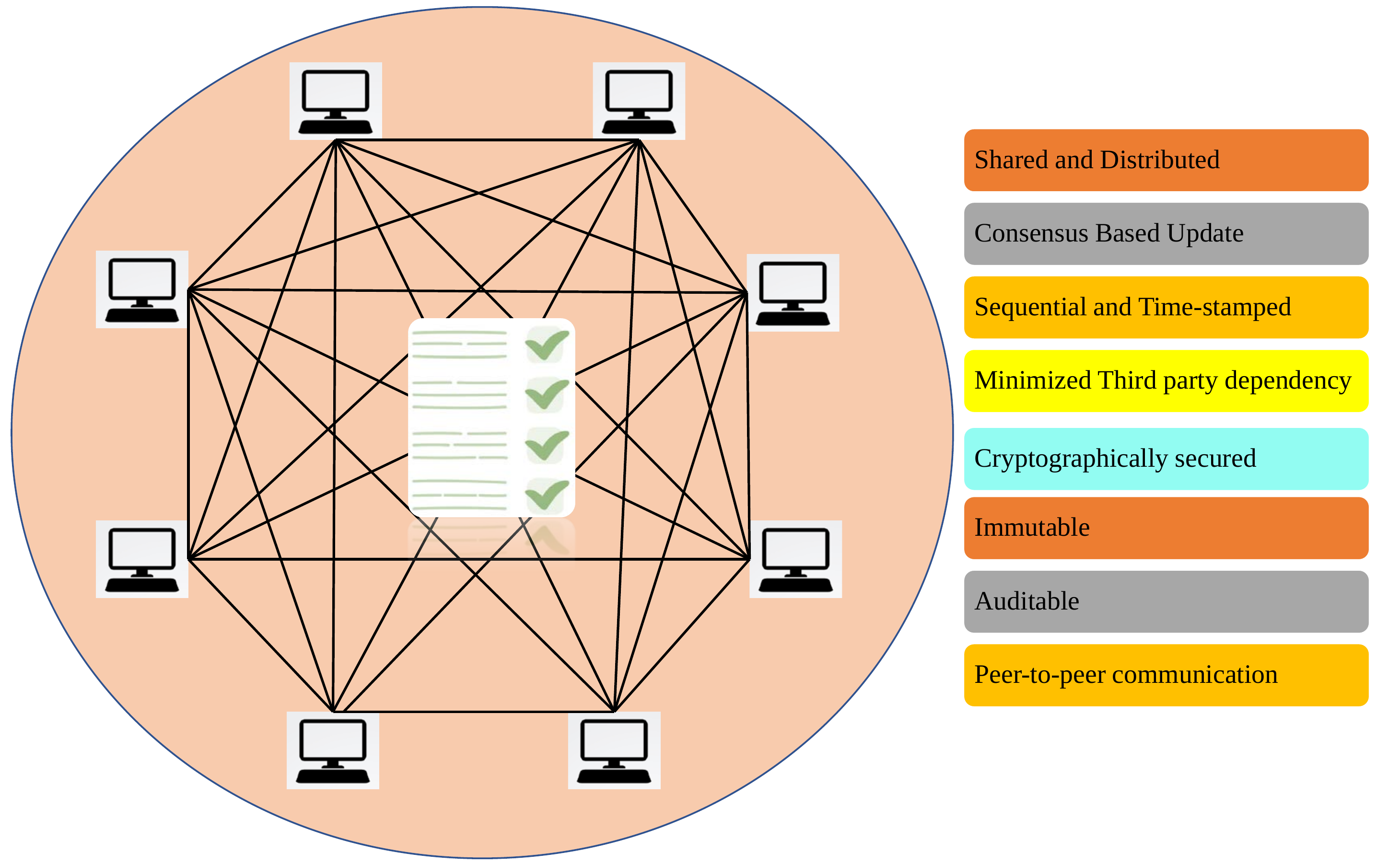}
	\caption{Characteristics of the Blockchain.}
	\label{fig:Characteristics_of_Blockchain}
\end{figure}

\subsubsection{Components of the Blockchain}

Block: The blockchain are chronologically connected blocks of hashed transactions \cite{Vallois_CSNet_2017}. Each block has a block header with all metadata apart from the transactions which are placed in the body of the block, as shown in Fig. \ref{fig:Header_of_Blockchain}. Transactions are separated from the metadata to increase the scalability of the blockchain by implementing pruning techniques. 

\begin{figure}[htbp]
	\centering
	\includegraphics[width=0.85\textwidth]{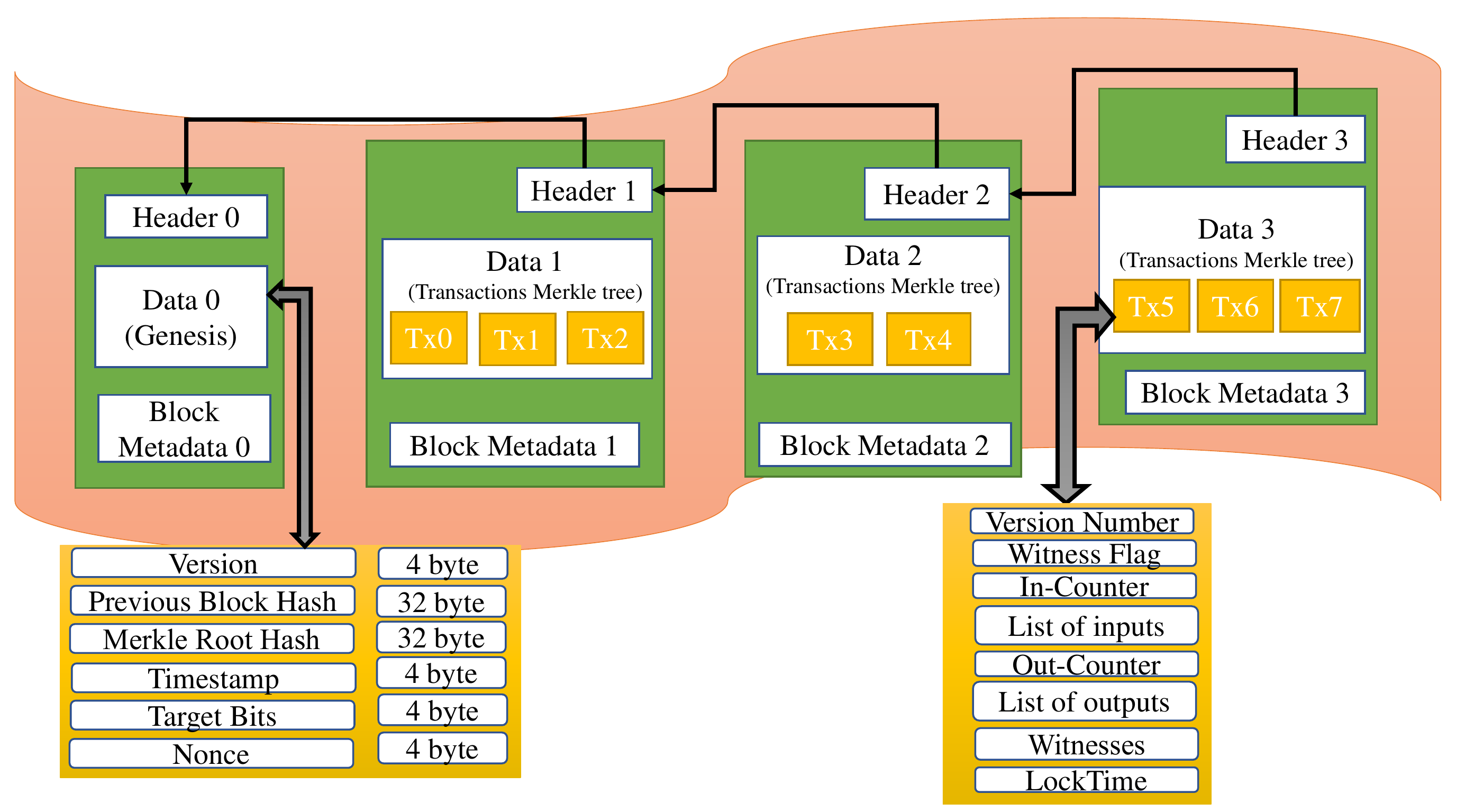}
	\caption{Header of the Blockchain.}
	\label{fig:Header_of_Blockchain}
\end{figure}

Structures may differ for different blockchains. Version represents the block version number and is typically updated when there are software upgrades. Previous block hash is the SHA-256 hash of the previous block header which creates a chain of blocks. All the transactions are hashed and connected as a Merkle tree. The root of the Merkle tree is hashed and the 256-bit hash is included in the block header. A timestamp is also included in the header which shows the current block creation time. Target bit indicate the difficulty of the cryptographic problem to be solved, and it varies based on the number of active peers and other parameters so as to maintain the block creation time to 10 minutes in bitcoin. A nonce is added to the block header which helps in Proof of Work (PoW )consensus mechanisms.

Node: All the participants in the blockchain network are called nodes. All the nodes can communicate with the blockchain to perform transactions. Every node will have its own copy of the ledger. Nodes storing all the transactions from the start of the first (genesis) block are called full nodes and can participate in the transaction verification process to check double spending on published transactions. Another special type of node are miners responsible for participating in the consensus mechanism and generating blocks from the transactions. Miners will receive incentives in terms of transaction fees and block rewards. Even though full nodes don’t have tangible incentives, their presence increases security. Full nodes require large amounts of storage space and miners require both space and computational power which demands special expensive hardware. 

Transaction: A typical bitcoin transaction represents a coin transfer between two peers, which are validated and grouped into blocks. These individual transactions can be generated by any of the peers and broadcast to the entire network. Each transaction will go through a validation and verification phase before adding it to a block. A full node is defined as a peer node with all the history of transactions in its memory from the start of the genesis block. Full nodes are responsible for transaction validations whereas miners are full nodes with high computational capabilities which participate in the consensus process and are responsible for building valid blocks from validated transactions. 

Consensus Mechanism: It plays a vital role in achieving agreement among the untrusted nodes participating in the network. It can be defined as a set of rules which are defined and accepted among all the nodes which will determine the conditions which should be satisfied for processing transactions and incentive information. This will also ensure consistency of the chain by defining some rules which help in resolving chain conflicts. Proof-of-Work (PoW) which was developed based on the hash cash algorithm, is used in bitcoin. PoW poses a computationally hard hashing problem to all the miners and the miner who will solve it first will be elected and will be responsible for creating the new block. These miners will be awarded with block reward and transaction fees in bitcoin. The PoW consensus mechanism is computationally and resource intensive which consumes a lot of power. As the blockchain has potential applications in a vast variety of fields and different fields have different requirements, different consensus mechanisms are invented for different applications to increase scalability. Some of them are Proof-of-Stake (PoS), Delegated Proof-of-Stake (DPoS), and Proof-of-Elapsed Time (PoET) \cite{Bodkhe_Access_2020}.  

\subsubsection{Network Operations in Blockchain} 

Described here is the sequence of operations performed from the time a transaction is published to the network until the block is generated and added to the ledger \cite{Puthal_CEM_2018}.

\begin{itemize}
	\item Multiple transactions are generated by the nodes participating in the network and all the transactions are broadcast to the network. It is not required for all transactions to reach all the participant nodes in the network. 
	\item Miner nodes in the network will group some of the transactions from the transaction pool. There is no need that all the miners will work on the same group of transactions. Generally, transactions with high transaction fee are likely to be picked by the miners because of the incentive they are going to be paid.
	\item Miners will change the nonce and compute the hash of the block header to achieve a hash below the targeted value which is a computational problem posed to the miners in PoW. 
	\item Once the hash is successfully computed, the block is broadcast by the corresponding miner.
	\item The transactions included in the broadcast block is verified by the nodes before adding it into the ledger. Once all the transactions are validated, incentives are allotted to the miner node.
	\item Other mining nodes will start the whole process again by excluding the transactions which are already confirmed in the previous block.
	\item The difficulty level of the PoW is adjusted based on the number of participants in the network and the amount of time defined for the block generation which is approximately 10 minutes in bitcoin. 
\end{itemize}

The overview of transaction processing is represented in Fig. \ref{fig:Overview_of_Transaction_Processing}. 

\begin{figure}[t]
	\centering
	\includegraphics[width=0.90\textwidth]{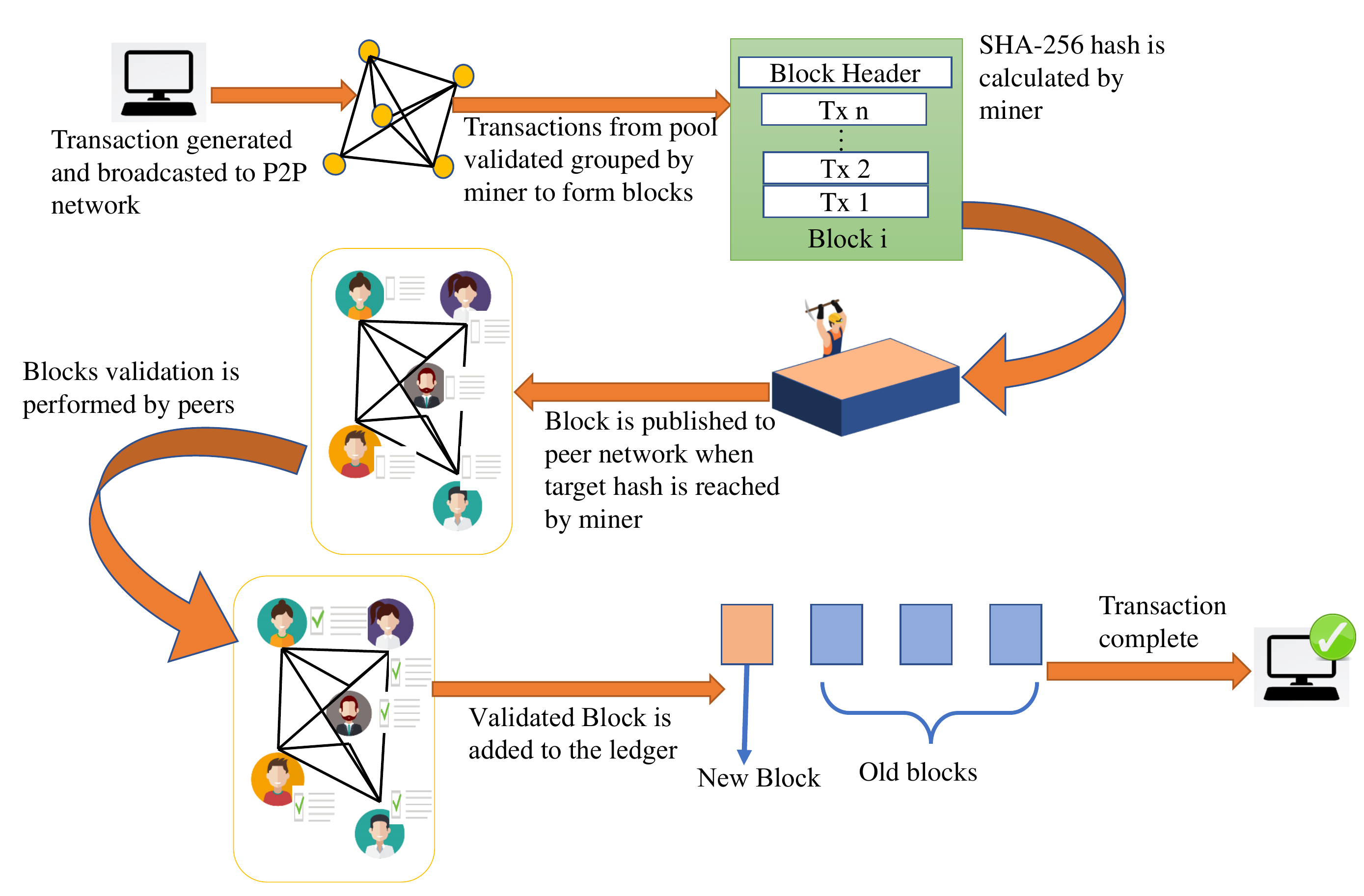}
	\caption{Overview of Transaction Processing.}
	\label{fig:Overview_of_Transaction_Processing}
\end{figure}

\subsection{Proposed Blockchain for Physiological Data Storage in SaYoPillow}
 
In the proposed architecture, we have assumed the network of sensors used for SaYoPillow is managed privately by the user. It is also assumed that each user will have their own personal private permissioned blockchain. Each blockchain will have an admin which is responsible for all the transactions of data and maintenance of the smart contracts. The proposed blockchain solution uses RSA encryption for communication between the edge device, end users and admin nodes. Transaction processing in the proposed blockchain solution can be divided into two different steps as discussed in this Section. The proposed blockchain architecture that is used in SaYoPillow is represented in Fig. \ref{fig:Proposed_Blockchain_Architecture_in_SaYoPillow}. 

\begin{figure}[htbp]
	\centering
	\includegraphics[width=0.95\textwidth]{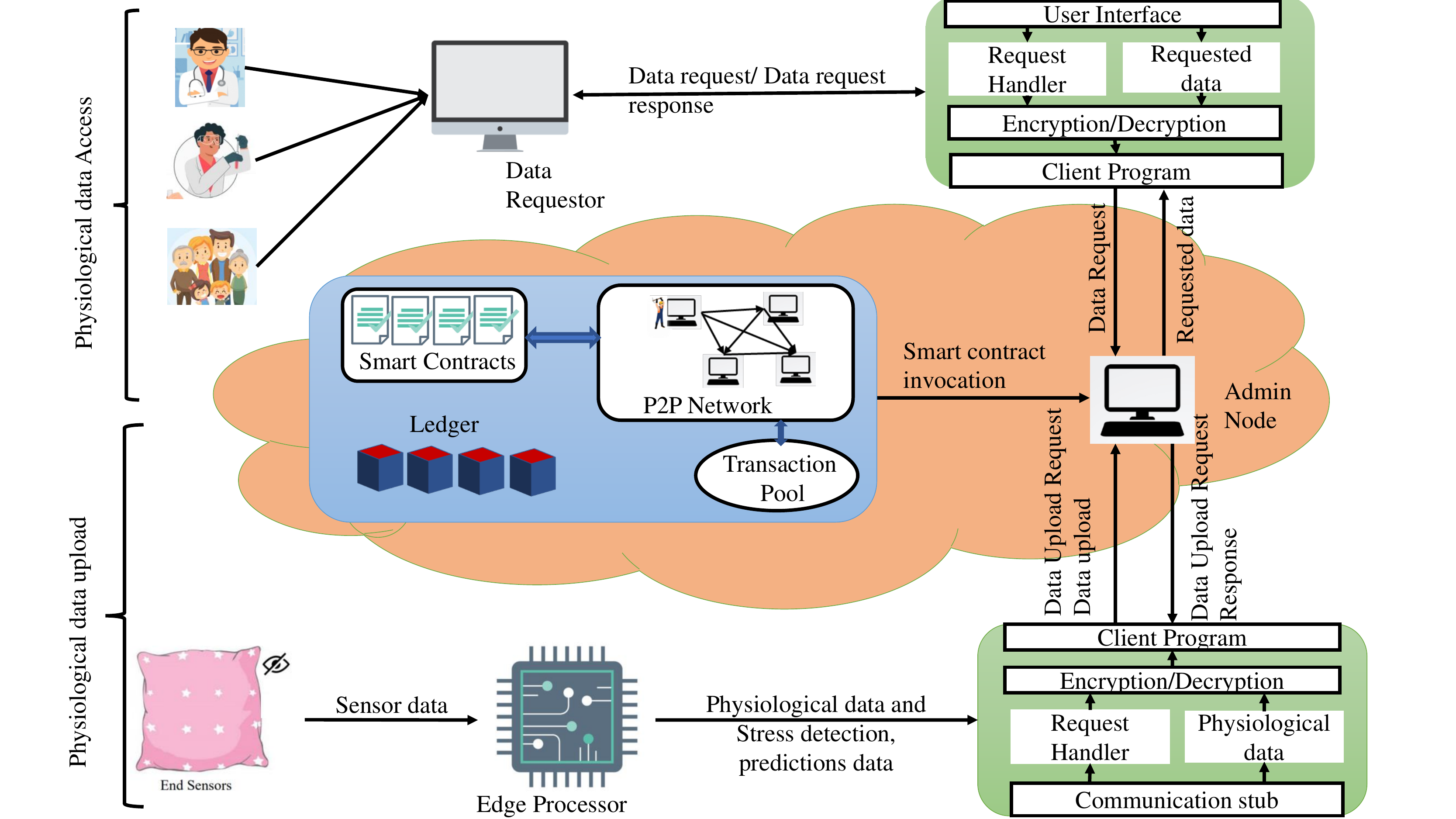}
	\caption{Proposed Blockchain Architecture in SaYoPillow.}
	\label{fig:Proposed_Blockchain_Architecture_in_SaYoPillow}
\end{figure}

\subsubsection{Data uploading from SaYoPillow}

Edge device in the SaYoPillow architecture is responsible for collecting physiological data which are used for detecting and predicting stress. The edge device creates a data upload request which will be sent to the admin node using a client program. The data upload request is strictly validated by the admin node using the access-policy smart contract. Once the user is validated, the admin node will acknowledge the request to the edge device which in turn uploads the gathered physiological and analyzed stress detection and prediction data. The data received by the admin node are published to the blockchain network. A smart contract is executed, and configured events are generated which will create logs in the blockchain ledger for auditing purposes. RSA encryption is used to securely transfer data between the blockchain and the edge device. The integrity of the requests is validated by using the digital signatures with the edge device private key and the public key of the admin node is used to maintain the confidentiality of the message in the communication channel. The process of uploading the data to the cloud from the edge device is detailed through Algorithm \ref{Algo5}. 

\begin{algorithm}[htbp]
	\begin{algorithmic}[1]
		\caption{Process of Uploading Gathered and Analyzed Data to Blockchain}
		\label{Algo5}
		\STATE Edge device and Admin node have their own Public and Private keys (PuE, PrE) and (PuA, PrA) respectively.
		\STATE A data upload request is generated by the Edge Processor. 
		\STATE Upload request is signed using the private key (PrE) of the edge device and signature is appended to the request. 
		\STATE Upload request along with signature is encrypted using the public key of the admin node (PuA) and published to admin node using client program.
		\STATE Decryption of the request and verification of the signature for message integrity is performed by Admin node. 
		\STATE Based on the public key, a strict access rights check is performed by using Access-policy smart contract. 
		\IF {Request is Authenticated}
		\STATE A response to upload the data is sent back to the Edge Processor.
		\STATE Another transaction with all the physiological data and detected and predicted information will be created by Edge Processor. 
		\STATE Before publishing to admin node, this transaction should be signed with private key (PrE) of the Edge Processor and should be encrypted using the public key (PuA) of the admin. 
		\STATE The digital signature should be checked to validate the integrity after receiving the transaction by Admin node.
		\IF {Signature is validated}
		\STATE Execute the upload function from data management smart contract. 
		\STATE Send back the Blockchain transaction and block information.
		\ELSE \STATE End the process.
		\ENDIF
		\ENDIF
		\STATE Repeat the steps from 4 through 19 every time there is a new data upload request. 
	\end{algorithmic}
\end{algorithm}

\subsubsection{Data retrieving from Blockchain}

End user prepares a transaction with the requested information. The transaction is encrypted using the public key of the admin node and it is sent using the client program. The client program will verify the user based on the access-policy list smart contract. When the user is authenticated, the requested data is pulled from the blockchain which is an encrypted message and is decrypted using admin private key. The decrypted information is then encrypted again with the requestor public key for securely transmitting over the channel. Requestor at the end will decrypt the received message using its private key to get the actual requested information. The detailed explanation of the process of data access from blockchain is explained through Algorithm \ref{Algo6}.

\begin{algorithm}[htbp]
	\begin{algorithmic}[1]
		\caption{Process of Accessing Gathered and Analyzed Data From Blockchain}
		\label{Algo6}
		\STATE Edge device and Admin node have their own Public and Private keys (PuE, PrE) and (PuA, PrA) respectively.
		\STATE A data access request is generated by the Requestor. 
		\STATE Data access request is signed by Requestor’s private key (PrR) and signature is appended along with request data.
		\STATE Data access request along with the signature is encrypted with Admin node public key (PuA) and published using client program.
		\STATE Admin node decrypts the received request and verifies the message integrity using the signature.
		\IF {There is a match in Siganture}
		\STATE Validate Requestor public key with access-policy smart contract.
		\IF {Requestor had Access Rights}
		\STATE Execute Data retrieve function of data management smart contract.
		\STATE Using Admin private key (PrA), decrypt the requested data.
		\STATE Using the Admin private key (PrA), add signature to the decrypted data.
		\STATE Using Pubic key of requestor (PuR), encrypt the requested data along with signature.
		\STATE Send this encrypted data to the requestor.
		\STATE Requestor should decrypt the received message using their private key (PrR) and verify the signature for data integrity.
		\STATE Decrypted information is the requested information.
		\ELSE 
		\STATE End the process of data access.
		\ENDIF 
		\ENDIF
		\STATE Repeat the steps from 4 through 19 every time there is a new data access request. 
	\end{algorithmic}
\end{algorithm}

\subsection{Proposed Approaches for Blockchain based Secure Storage and Access}

As sensitive personal information is shared in SaYoPillow, security is a major concern. In this section we have taken different attack scenarios to evaluate the efficiency of our proposed model to mitigate the risks.

\textbf{Threat 1:} A scenario where an adversary is trying to imitate a user of the SaYoPillow and trying to upload manipulated physiological data. \\
\textbf{Solution 1:} While uploading physiological data, the edge device must sign the data with its own private key. The digital signature is appended with the physiological data and the whole transaction is encrypted with the public key of the admin node. The admin node will decrypt the transaction request with its private key and check the signature for integrity. The private key is never transmitted over the network. The adversary will not have access to the private key of the edge device which makes it impossible for sending an authorized transaction. Hence, this threat scenario can be blocked by using the proposed model.

\textbf{Threat 2:} A threat scenario where an adversary is trying to make an unauthorized retrieval of information from the private blockchain. \\
\textbf{Solution 2:} Whenever a data access request is made to the admin, the admin node will verify the public key of the requester against the strict access-policy smart contract. The requested data are retrieved only when the public key of the requester is assigned with a role which has access rights by the user of SaYoPillow. As the adversary public key will not be in the access-policy list of the smart contract, no data will be retrieved or displayed. 

\textbf{Threat 3:} A eavesdropping scenario is considered where an adversary is trying to listen to the network traffic while an authorized node is retrieving the physiological information from the blockchain. \\
\textbf{Solution 3:} After the admin node processes the request from an authorized requester, the retrieved data are again encrypted using the public key of the requester before sending it into the network channel. As the data is encrypted with the public key of the requester, only the requester which has access to its own private key can decrypt them. Even if the adversary, who does not have access to the private key of the requester, intercepts the messages still cannot make any sense from it. This way the proposed model can avoid unauthorized access of data.

\textbf{Threat 4:} A scenario where an adversary has gained access to the cloud environment in which the private blockchain is running. \\
\textbf{Solution 4:} While uploading the physiological data and analyzed stress data, the digital signature of the edge device is encrypted with the admin node public key before sending data to the admin. The admin node will retrieve the information and before publishing the data to the blockchain, the data are encrypted using the admin public key. The data in blockchain are stored as an SHA-256 digest, hence the adversary will not be able to understand the information present in the cloud environment even if able to access it.

\section{Implementation and Validation of SaYoPillow}
\label{SEC:SaYoPillow_Implementation}
  
\subsection{Implementation of SaYoPillow using Off-The-Shelf Components}

The physiological signal data at the user end are taken and processed by the edge device. The 15,000 sample dataset is fed in to the device for processing and the stress state detection is performed. The stress prediction for the following day is performed after the person is completely awake. The data are securely sent continuously to the cloud storage from where the stress levels detected and predicted along with the average physiological signal data are presented in the UI. The stress control mechanisms during sleep are assumed to be controlled automatically by the edge device and thus the usage of 5 LED lights for each stress state and appropriate actions as represented in Fig. \ref{fig:Implementation_Setup}.  

\begin{figure}[htbp]
	\centering
	\includegraphics[width=0.85\textwidth]{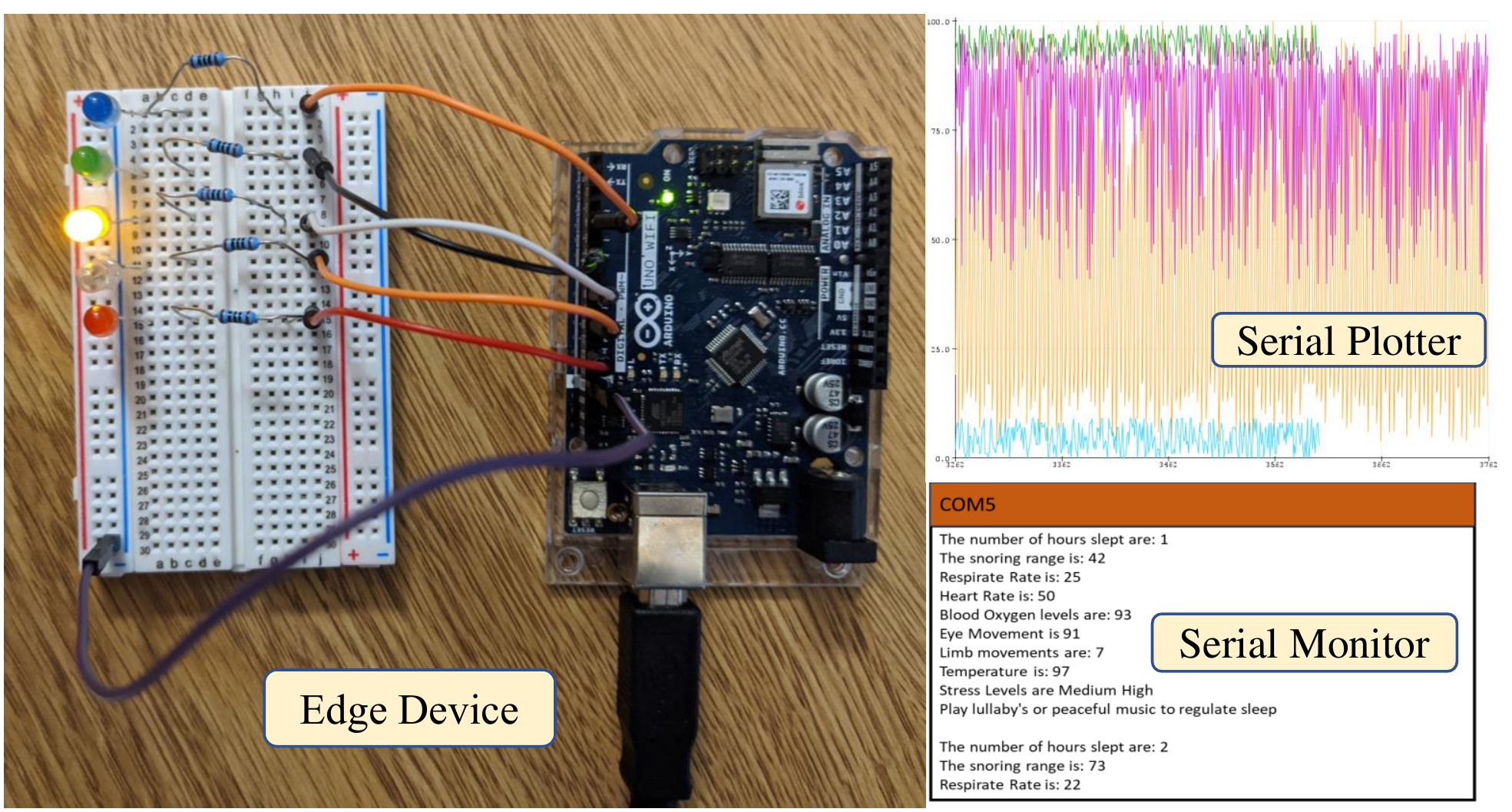}
	\caption{Implementation Setup in SaYoPillow.}
	\label{fig:Implementation_Setup}
\end{figure}

Once after the data is sent from the edge device, the secure cloud storage and access is provided in SaYoPillow by implementing an Ethereum blockchain. Ethereum is an open-source blockchain based platform that provides a blockchain with a built-in Turing-complete language which can be used to program smart-contracts \cite{Ethereum_Whitepaper}. In SaYoPillow each user is assumed to have their own private blockchain through which the data is shared among other entities using a strict access control policy. 

Ethereum blockchain stores all the transactions in ledgers, with each node participating in the network having its own copy of the ledger. It stores the state of each smart contract in the network and updates it every time a transaction is performed.
Ethereum uses accounts to keep track of the user’s balance. Ethereum provides a programmable environment which provides the Ethereum Virtual Machine (EVM) for reading and executing byte codes of smart contracts. A smart contract can be defined as set of rules described by the developer to enforce between users using cryptographic codes when certain conditions are met. Bitcoin to some extent supports smart contracts by providing few hundreds of scripts, most of which can be applicable to financial transactions. Ethereum on the other hand replaces the scripts by defining a language and supports a wide range of operations. Through the EVM executing bytecode, developers can write smart contracts in a high-level language called solidity \cite{solidity_language}. Every time a new transaction is created for smart contract, a network of participants will execute the smart contract and come to an agreement. Smart contracts can also store application information. This type of processing and achieving consensus will remove the necessity of  central authorities and will help in leveraging decentralized applications which can run on P2P networks.   

For SaYoPillow, a private Ethereum blockchain has been created using a cloud platform and three Linux/Unix Elastic Cloud Compute (EC2) services and its deployment, status and explorer are represented in Fig. \ref{fig:Ethereum Blockchain in SaYoPillow.}.  All three instances are of type t2.medium with 2 virtual CPUs attached. A separate t2.micro type is used to develop an admin node in the proposed model. A Geth client (Go-Ethereum implemented in Go language) is required for the end devices to interact with the blockchain smart contracts. A Geth client is installed at the edge device in order to send physiological and stress detected, stress predicted and control  information to the blockchain and another Geth client is installed at the end user for retrieving requested information from the private blockchain.

\begin{figure}[htbp] 
	\centering
	\subfigure[Ethereum Blockchain Deployment]
	{\includegraphics[width=0.75\textwidth]{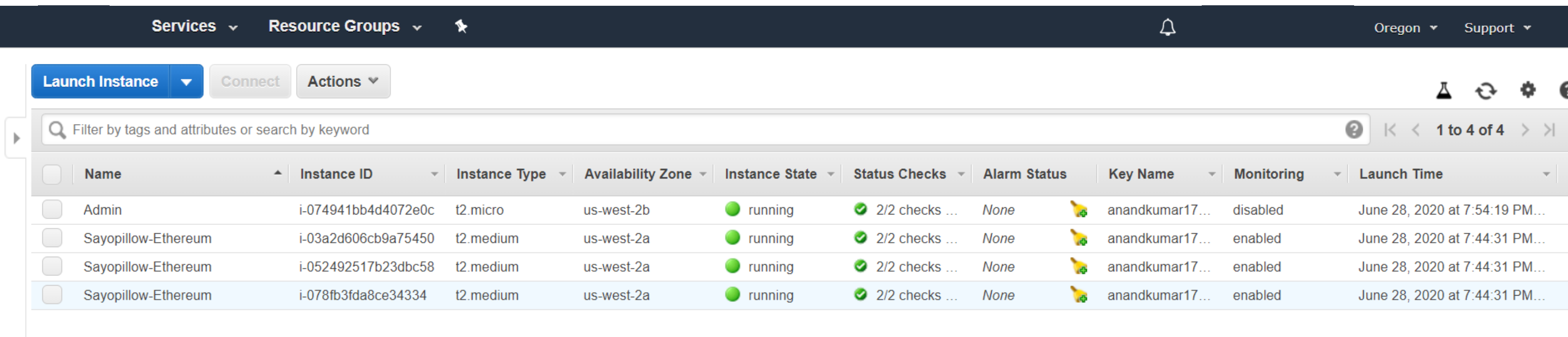}}	
	\subfigure[Ethereum Blockchain Status]
	{\includegraphics[width=0.75\textwidth]{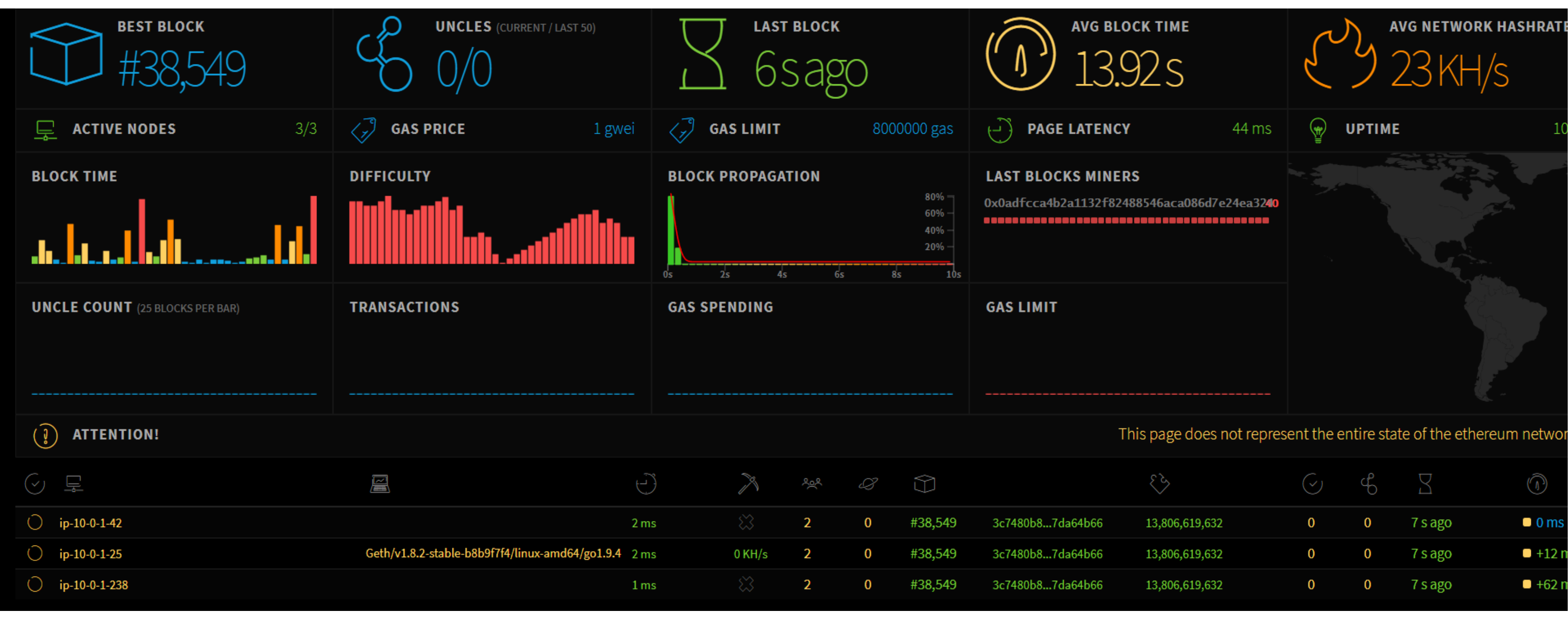}}
	\subfigure[Ethereum Blockchain Explorer]
	{\includegraphics[width=0.75\textwidth]{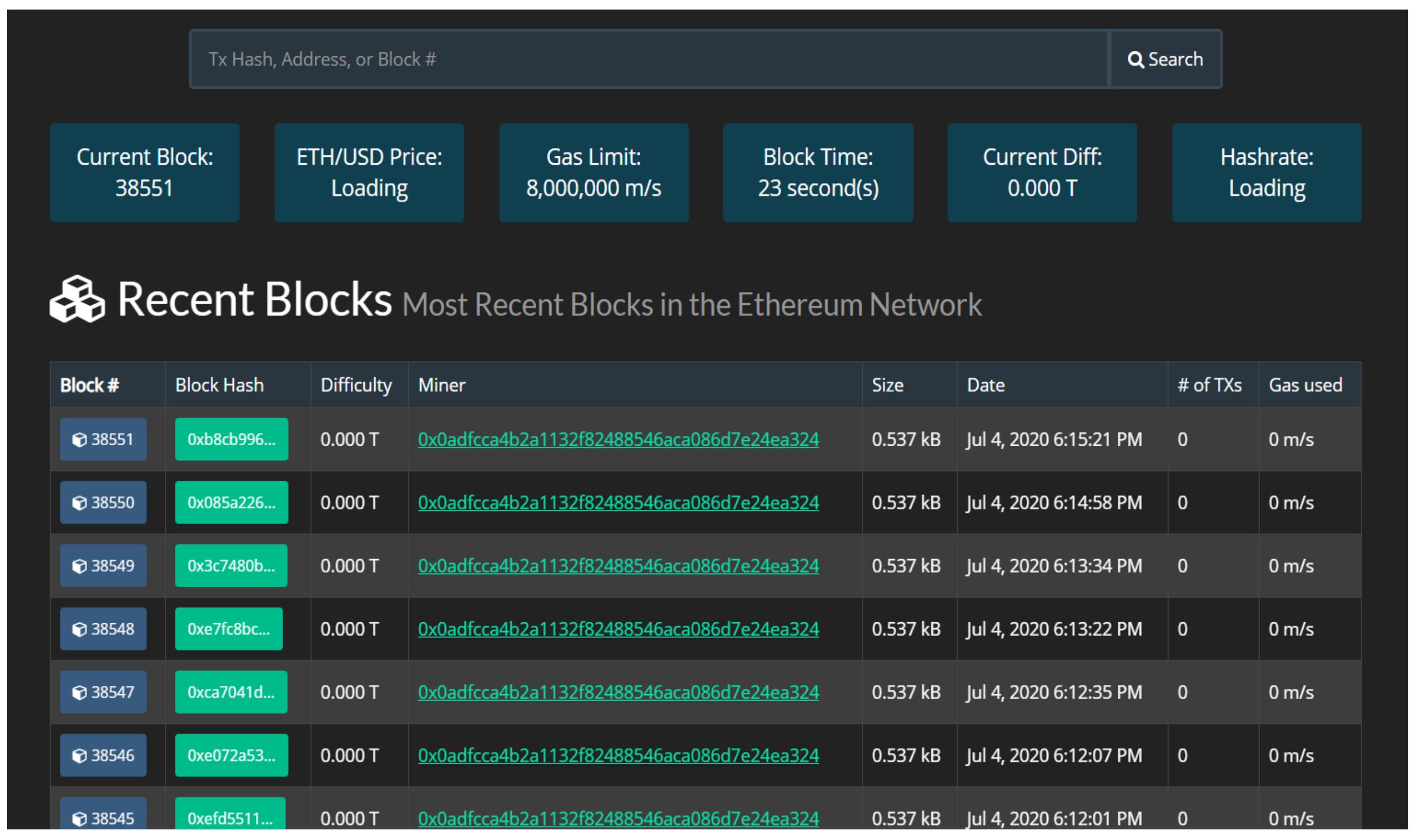}}
	\caption{Ethereum Blockchain in SaYoPillow.}
	\label{fig:Ethereum Blockchain in SaYoPillow.}
\end{figure}

Two different smart contracts have been defined for access control and data management in SaYoPillow. An access control management smart contract provides capabilities of adding different roles and assigning roles to different entities by the user.
The user can control by whom the data can be accessed to this smart contract, thus maintaining a strict access control policy. Similarly, the user also has control to remove any added roles or role bearers. All these operations in this smart contract will create auditable logs which are stored in the blockchain and are immutable.

A User Interface is built using html and web3.js \cite{web3j} scripts to interact with the private Ethereum blockchain. The metamask browser extension \cite{MetaMask} is used to enable users to access Ethereum web sites . Along with the transaction creation, the UI representation with two different users for accessibility testing, one of which is added to a ``Family'' role and the other is not assigned any access permissions is shown in Fig. \ref{fig:Ethereum Blockchain Process and UI in SaYoPillow.}. Stress control  mechanisms are also provided in the UI for the user.

\begin{figure}[htbp] 
	\centering
	\subfigure[Creating a Transaction in Ethereum]
	{\includegraphics[width=0.65\textwidth]{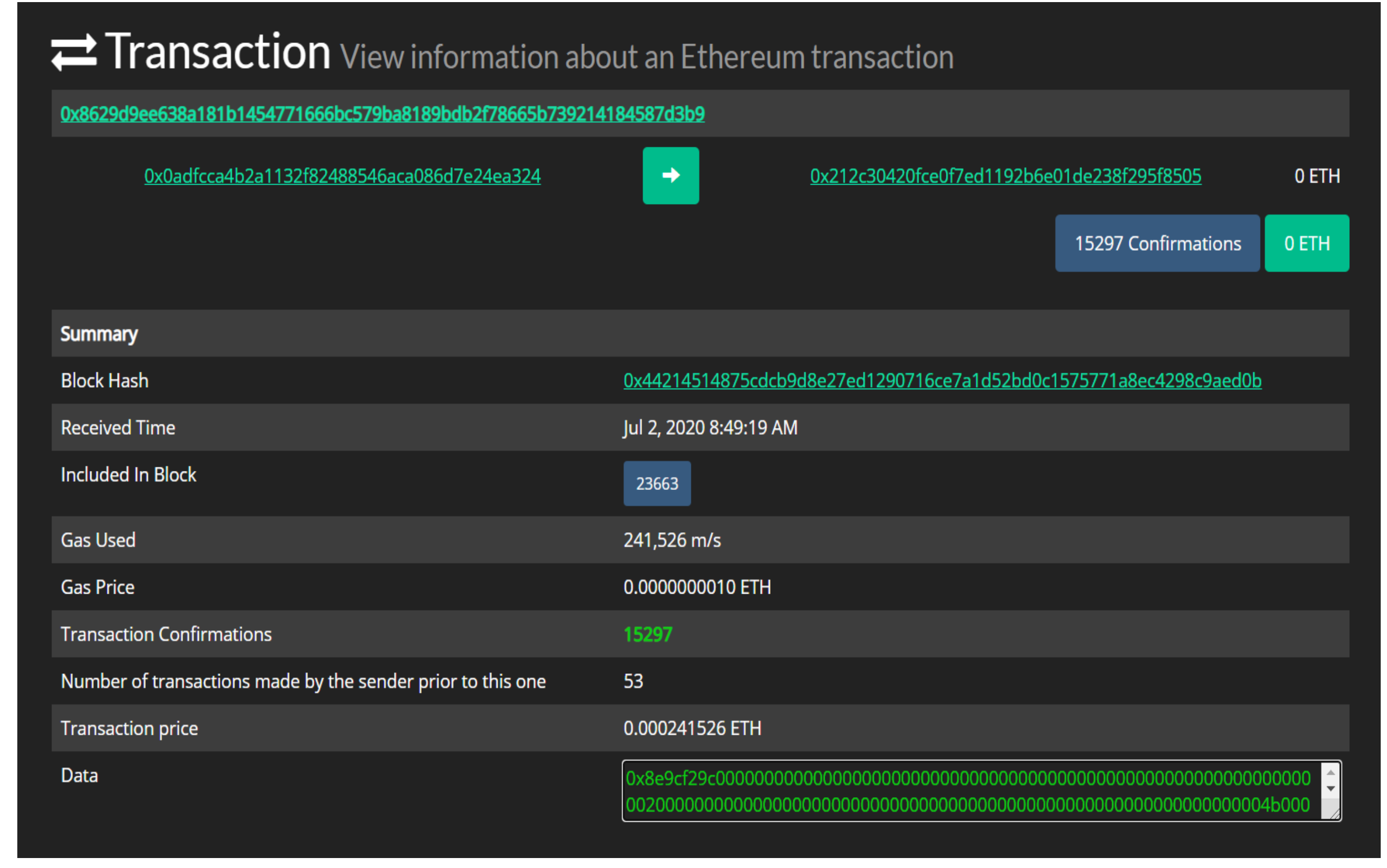}}	
	\subfigure[UI with Access]
	{\includegraphics[width=0.65\textwidth]{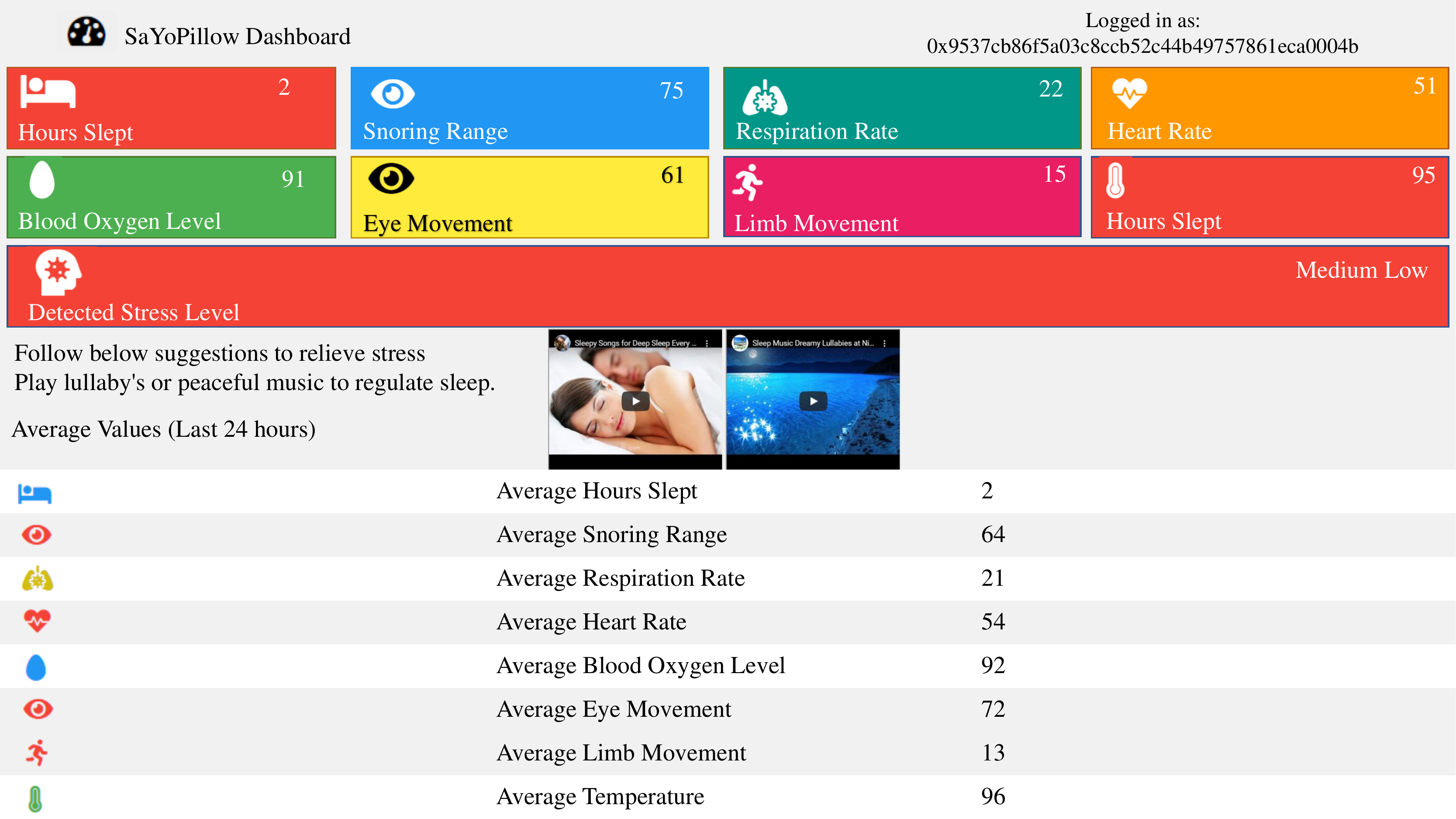}}
	\subfigure[UI without Access]
	{\includegraphics[width=0.65\textwidth]{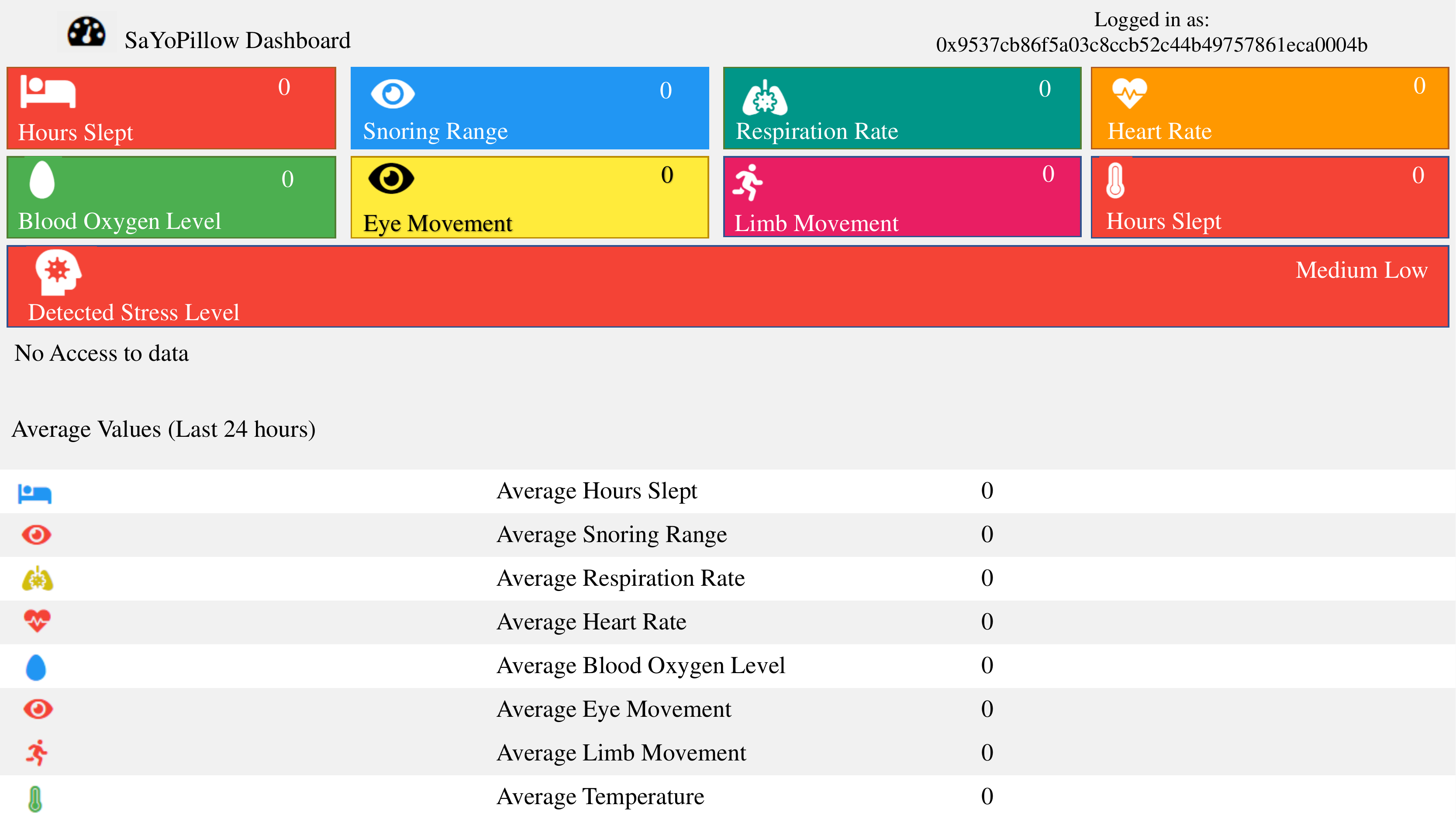}}
	\caption{Ethereum Blockchain Process and UI in SaYoPillow.}
	\label{fig:Ethereum Blockchain Process and UI in SaYoPillow.}
\end{figure}

\subsection{Validation of SaYoPillow}

\subsubsection{Validation of Stress Monitoring}
For the analysis of stress detection and prediction, an edge device at the user end is used to process the physiological signal data. The implementation was performed using TensorFlow as 15,000 samples of data were used to test and train the model. The analyzed stress levels are sent securely to the cloud for storage. The accuracy is approximately 96\% for the model to detect the stress as shown in Fig. \ref{fig:Loss_and_Accuracy_of_SaYoPillow}. The stress prediction is performed at the UI by taking in the total stress levels detected every 15 minutes and comparing the observed stress levels, as detailed in Algorithm \ref{Algo4}. The characteristics of SaYoPillow are represented in Table \ref{Table: Characteristics of SaYoPillow}.

\begin{table}[htbp]
	\caption{Characteristics of SaYoPillow}
	\label{Table: Characteristics of SaYoPillow}
	\centering
	\begin{tabular}{|p{4.0cm}p{3.6cm}|}
		\hline \hline
		\textbf{Characteristics} & \textbf{Specifics}\\
		\hline 
		\hline
		Data Acquisition & NSRR Sleep Study Dataset \\
		\\
		Data Analysis Tool & TensorFlow Lite \\
		\\
		Input Dataset & 15000 samples \\
		\\
		Classifier & FCNN \\
		\\
		Stress level Classification & 5 \\
		\\
		Accuracy & 96\%  \\
		\hline
		\hline
	\end{tabular}
\end{table}

For any classification machine learning model the Precision and Recall curves are important to validate its effectiveness. The Precision Vs Recall curve is represented in Fig. \ref{fig:Precision_Recall}. 

\begin{figure}[htbp]
	\centering
	\includegraphics[width=0.65\textwidth]{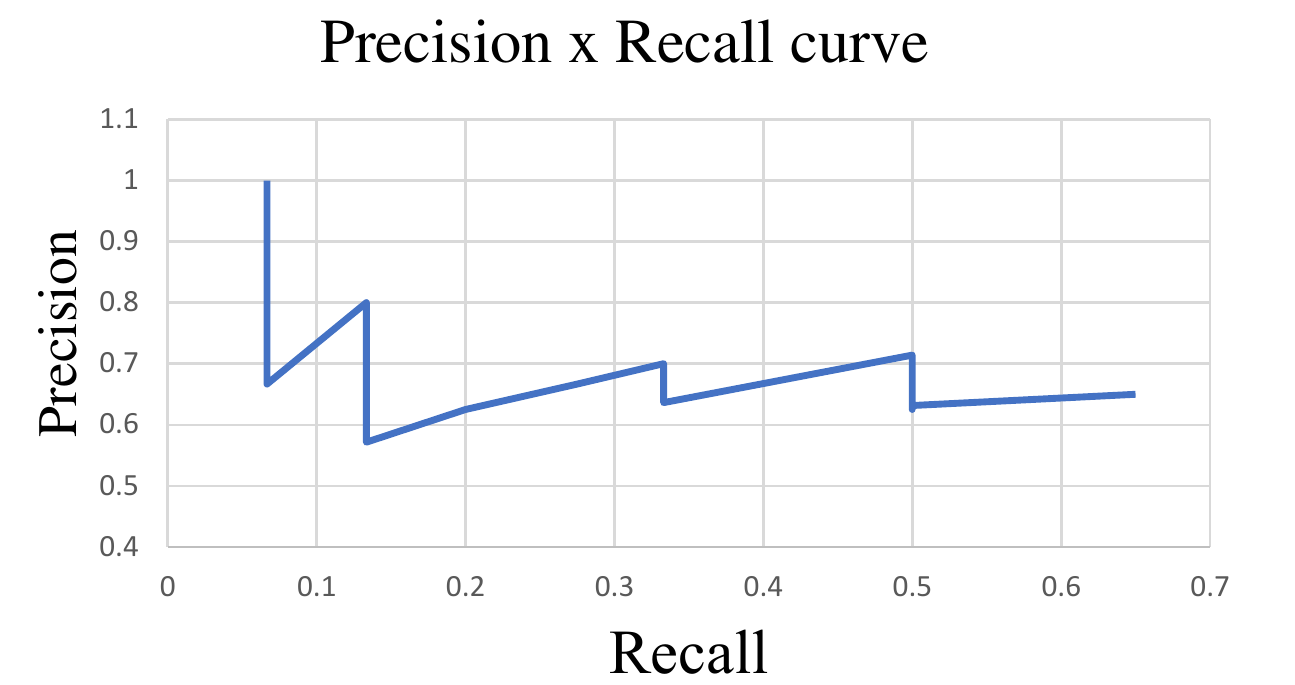}
	\caption{Precision Vs Recall in SaYoPillow.}
	\label{fig:Precision_Recall}
\end{figure}

\subsubsection{Validation of Stress Control}
For the stress control during sleep, the edge device is programmed in such a way that it controls the ambiance of the room to regulate sleep by following remedies as explained in Sec \ref{Section:Stress and Sleep} and Sec. \ref{Sec. Stress Control}. Maintaining the ambiance of the room by regulating temperature, dimness of lights, playing lullabies, releasing sleep regulating essential oils are proposed in SaYoPillow. The control remedies for next day stress prediction are provided in the UI for the user. 

\subsubsection{Validation of Secure Storage and Access}


A data management smart contract is created for storing and retrieving physiological data from the edge device. The three main important functions performed here are createPhysiologicalRecord, which can only be executed by the user, takes analyzed stress levels along with the physiological data from the edge device and stores it in smart contract in-memory mappings, retrieveLatestRecord retrieves the latest information captured by the edge device and presents it to the requester, and averageValues presents the overall evaluation of stress and other physiological parameters for 24 hours. The retrieveLatestRecord function interacts with the access control smart contract to check the access policies before retrieving the physiological information. 

For the validation, evaluation of a few metrics is performed in both the Ropsten test network (public) and our SaYoPillow (private) deployed in the cloud. Ropsten is a test network provided to simulate transactions like production Mainnet Ethereum. Transaction Time (TT) is taken as a parameter for evaluating the performance. Transaction times are calculated from the time of submitting the transaction to getting the receipts for 10 transactions for each function. Average amount transaction times for individual functions are calculated by taking the mean of all recorded transaction times, as shown in Table \ref{Table: Evaluation Metrics of SaYoPillow}.

\begin{table*}[htbp]
	\caption{ Metric Evaluation for SaYoPillow}
	\label{Table: Evaluation Metrics of SaYoPillow}
	\centering
	\begin{tabular}{p{1cm} p{3cm} p{3cm} p{3cm} p{3cm} p{3cm}|}
		\hline \hline
\textbf{Network} & \textbf{Contract Deployment: Min, Max and Avg TT (secs)} & \textbf{Adding Role: Min, Max and Avg TT (secs) } & \textbf{Adding Role Bearer: Min, Max and Avg TT (secs)} & \textbf{Creating Data Record: Min, Max and Avg TT (secs)} \\
\hline
\hline
\\
		Ropsten & \begin{tabular}{p{0.5cm}|p{0.5cm}|p{0.5cm}} 3.29 & 26.75 & 11.8 \\ \end{tabular} 
				& \begin{tabular}{p{0.5cm}|p{0.5cm}|p{0.5cm}} 1.2 & 18.4 & 8.6 \\ \end{tabular}
			 	& \begin{tabular}{p{0.5cm}|p{0.5cm}|p{0.5cm}} 1.4 & 35 & 15 \\ \end{tabular}
			 	& \begin{tabular}{p{0.5cm}|p{0.5cm}|p{0.5cm}} 1.5 & 38.2 & 11.2\\ \end{tabular} \\
		\\		
		SaYoPillow & \begin{tabular}{p{0.5cm}|p{0.5cm}|p{0.5cm}} 3.2 & 13.5 & 6.3 \\ \end{tabular} 
		& \begin{tabular}{p{0.5cm}|p{0.5cm}|p{0.5cm}} 1.4 & 10.7 & 5.4 \\ \end{tabular}
		& \begin{tabular}{p{0.5cm}|p{0.5cm}|p{0.5cm}} 1.5 & 14.2 & 5.4 \\ \end{tabular}
		& \begin{tabular}{p{0.5cm}|p{0.5cm}|p{0.5cm}} 2.2 & 11.5 & 8.9\\ \end{tabular} \\
		\hline
		\hline
	\end{tabular}
\end{table*}

An overview of the evaluations can be seen in the Fig. \ref{fig:Ropsten_vs_SaYoPillow_Graph}. Transaction times mainly depend on the number of unconfirmed transactions in the transaction pool and the number of miners in the network. As we can see mean transaction times in private instances for all functions are lower than mean transaction times in the test network Ropsten. This is because of the unconfirmed transaction pool, as Ropsten is a public test network and has more transactions in the pool compared to private instances. The gas limit is maintained constant in both networks as transaction times will vary based on the offered gas price. SaYoPillow is configured to update the physiological data in intervals of 15 minutes, hence the number of access requests is very high compared to other data upload functions. This is an advantage as data access uses call methods and is retrieved in near real-time.  Private instances are preferred in our implementation to enhance security by limiting the participants in the network to only authorized. One more advantage with a private Ethereum network is that the difficulty of mining can be set to acceptable values based on the application, to increase scalability and reduce stalling of transactions in an unconfirmed state for long time.   

\begin{figure}[htbp]
	\centering
	\includegraphics[width=0.75\textwidth]{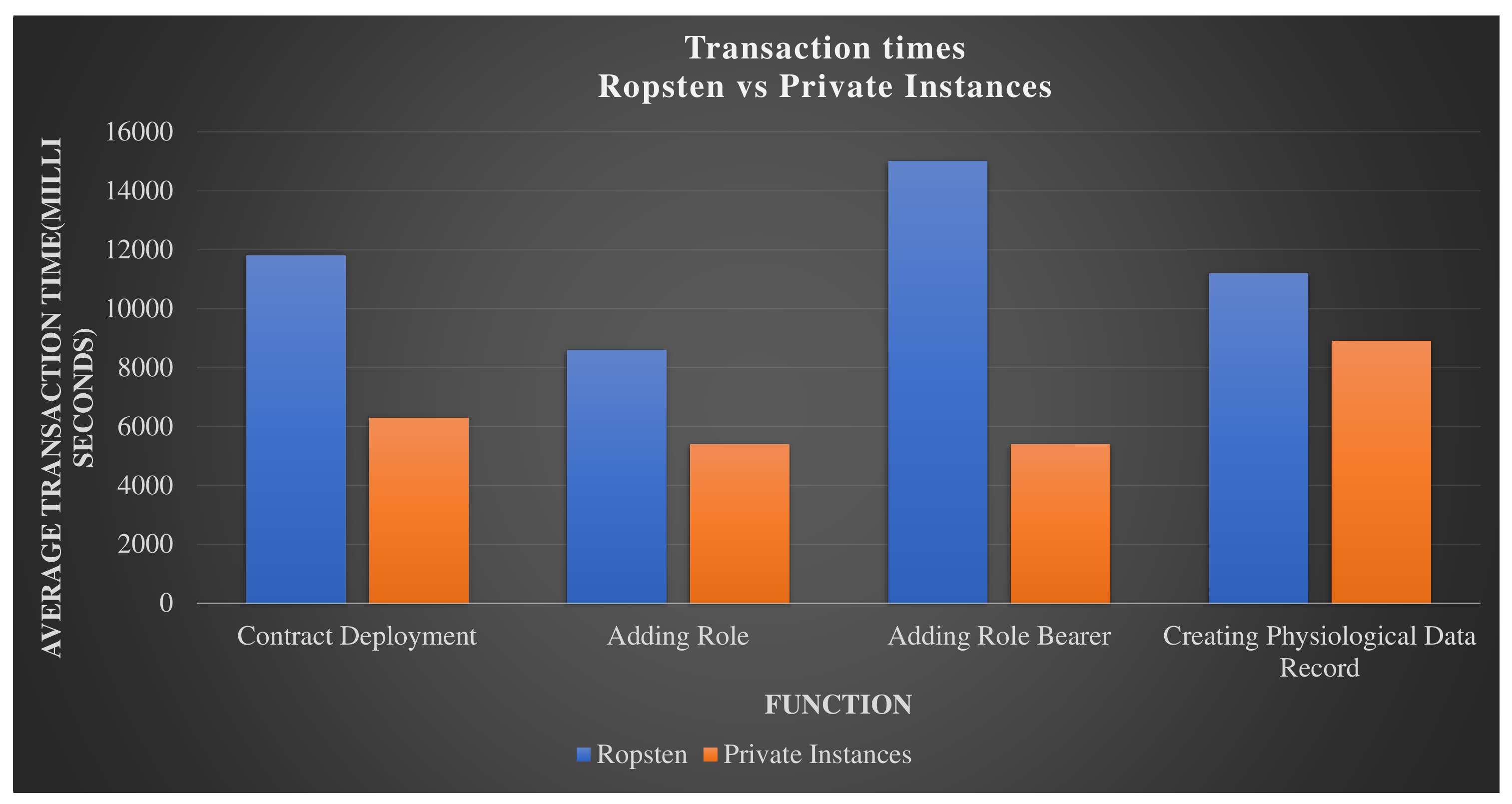}
	\caption{Ropsten Vs SaYoPillow Network Comparison Graph.}
	\label{fig:Ropsten_vs_SaYoPillow_Graph}
\end{figure}

\subsection{Comparison of SaYoPillow with State-of-the-art}

Though there are many state-of-the-art works along with the products that are available in the market, as discussed in \ref{SEC:Related_Research}, SaYoPillow has a potential to be marketed as it produces very efficient results in terms of classifying the monitored, detected and predicted stress levels to five states, providing stress control mechanisms, and providing a secure data transfer and storage. A brief comparison of SaYoPillow to the existing literature is presented in Table \ref{Table: Comparative Analysis}.

\begin{table*}[t]
	\caption{Comparative Analysis of Sleep Quality Monitoring Systems}
	\label{Table: Comparative Analysis}
	\centering
	\begin{tabular}{|p{1.7cm}p{2.6cm}p{1.5cm}p{1.0cm}p{1.0cm}p{2.1cm}p{1.0cm}p{1.0cm}p{1.0cm}|}
		\hline \hline
\textbf{Research Works} &	\textbf{Stressors} & \textbf{ML Algorithm} & \textbf{Features Extracted} & \textbf{Stress Levels} & \textbf{Factors Considered}  & \textbf{Stress Control} & \textbf{Security Feature} & \textbf{Accuracy (\%)} \\
		\hline
		\hline 
		Ciabattoni, et al. \cite{Ciabattoni_ICCE_2017} & GSR, RR, BT & KNN & 10 & 2 & Cognitive Tasks & No & No & 84.5 \\
		\\
		Rachakonda, et al. \cite{Rachakonda_TCE_2020} & Images & SSD Mobilenet & 4 & 2 &  Food Consumption & Yes & No & 98\\
	    \\
		Lawanot, et al. \cite{Lawanot_ICCE_2019} & Images and Surveys & KNN, SVM, DT & 12 & 2 & Regular Activities & No & No & 83 \\
		\\
		Sunthad, et al. \cite{Sunthad_ICCE_2019} & NIR & RNN & 4 & 3 & Classroom Activities & No & No & N/A \\
		\\
		Nath, et al. \cite{Nath_ICCE_2020} & GSR, PPG & RCM & 17 & 2 & Cortisol Levels & No & No & 92 \\
		\\
		Rachakonda, et al. \cite{Rachakonda_TCE_2019} & Body Temperature, Steps taken and Humidity & DNN & 3 & 3 & Physical Activities & Yes & No & 98.3 \\ 
		\\
		Ciabattoni, et al. \cite{Ciabattoni_ICCE_2020} & Mouse and Keyboard usage & KNN, DT, SVM & 62 & 2 & Work Activities & No & No & N/A \\
		\\
		 Wu, et al. \cite{Wu_JBHI_2019} & HRV & KNN & N/A  & 2 & Self Tracking & No & No & 97 \\
		\\
		Arsalan, et al. \cite{Arsalan_JBHI_2019} & EEG and fNIRS  & SVM, NB, MLP & 5 & 2 and 3 & Physical Activities & No & No & 92.85 and 64.28 \\
		\\
	 Mitrpanont, et al. \cite{Mitrpanont_ICT_2017} & EEG & N/A & 6 & 2 & Physical Activities & No & No & N/A  \\
		\\
		Nil, et al. \cite{Gurel_SJ_2019} & ECG and SCG  & LOSO-CV & 6 & 2 & Physical Activities & No & No &85 \\ 
		\\
		Rachakonda, et al. \cite{Rachakonda_ISES_2018} & Body Temperature, Heart Rate, Snoring range and Sleep Duration & N/A & 4 & 5 & Sleeping Habits & Yes & No & N/A \\
		\\
		\textbf{SaYoPillow (Current Paper)} & PSG, Snoring Range, Sleep Duration  & FCNN & 8 & 5 & Sleeping Habits & Yes & Yes & 96 \\
		\hline
		\hline
	\end{tabular}
\end{table*}

\section{Conclusions and Future Research}
\label{SEC:Conclusion}

Along with the stress monitoring and analysis during sleep, the importance of Smart-Sleeping is made understood to the user in SaYoPillow. The continuous monitoring of eight physiological signal data is processed in an edge device where five different classifications of stress is performed. The stress during sleep and stress prediction for the following next day are observed to be approximately 96\% accurate by implementing a Fully Connected Neural Network Model using TensorFlow. The dataset size which was used is 15,000 samples, out of which 13,000 are used for training while 2,000 samples are used for testing. The loss is observed to be less than 1\%. The fuzzy logic GUI has also been represented with a surface plot to allow the idea of extending this to develop a mobile application. 

The edge level implementation is done by feeding the device with the dataset and observing the analyzed stress state along with the instant physiological signal data in the monitor. This device was connected via Wi-Fi to directly send the analyzed information securely to the cloud for storage. The Ethereum blockchain was used as a cloud for storing the analyzed data. Every time the user received it, it is stored and analyzed information is displayed in the connected User Interface when the person is awake. The UI displays the average of the physiological signals for every 15 minutes throughout the sleeping period including the sleep latency periods, along with the detected stress level and its remedies for the user. The predicted stress level of the person along with its remedies are also displayed here.

The UI being built interacting with Ethereum blockchain allows the user to add or remove roles thereby providing access controls. The UI was built on an end user device, which has a great potential to be developed in to a mobile application for future research. Also, third party applications or existing mobile applications can be interfaced to the existing SaYoPillow for more versatile applications. SaYoPillow, an individual device promises to monitor stress and secure data storage with no compromise of safety or privacy of the users, could be a potential marketable device as it is reliable.

The current SaYoPillow provides a significant security and privacy aware smart healthcare framework which can drive research in many directions \cite{Sundaravadivel_MCE_2018-Jan}. It can have future research in terms of various other healthcare applications, such as diet monitoring and control, glucose-level monitoring and control, and seizure detection and control. It can also drive various other security and privacy issues that are critical using blockchain and physical unclonable function (PUF) like disruptive mechanisms \cite{Mohanty_MCE_2020-Mar}. 

The work can be further extended in the specific domain of stress monitoring and control as it is an important problem that can have significant social impact. The studies which relate stress to sleeping habits, sleep impairment and physiological/psychological problems which can be recorded during a crisis or after a crisis can have significant impact. SaYoPillow be developed as a device which can generate longitudinal data, i.e. data related to stress and sleep of individual subjects over a long period of time. This can help in analyzing the stressors \cite{Rachakonda_ISES_2018, Akerstebt_2006}. This stressors data could be used as biomarkers for future diagnosis including severe health hazards like insomnia, depression, cardiovascular disorders, migraine, cancer and lung disorders. Further improvements in SaYoPillow can make a huge difference in the medical world by helping reduce drug prescriptions by regulating stress and can help in achieving normal stress response.

On the security, privacy, and access control, many possible forms of research can be under taken \cite{Mohanty_arXiv_2019-Sep17-1909-06496_PUFchain, Biswas_Computer_2020-Jul, Puthal_Potentials_2019-Jan}. Blockchain based solutions have their energy and resource overheads. While blockchain solutions can be effective for the IoMT-cloud, it is still a quite challenging task to have a blockchain based solution for IoMT-edge. Further more, it is much more difficult to have blockchain based solution for IoMT-end devices involving implantable and wearable medical devices (IWMDs).

\section{Acknowledgments}

This material is based upon work supported by the the National Science Foundation under Grant number OAC-1924112. Any opinions, findings, and conclusions or recommendations expressed in this  material are those of the author(s) and do not necessarily reflect the views of the National Science Foundation.

The authors acknowledge inputs from Dr. Madhavi Ganapathiraju and Ms. Kalyani Karunakaran during initial versions of this work.

\bibliographystyle{IEEEtran}



\vspace{-1.0cm}

\begin{IEEEbiography}
[{\includegraphics[height=1.3in,keepaspectratio]{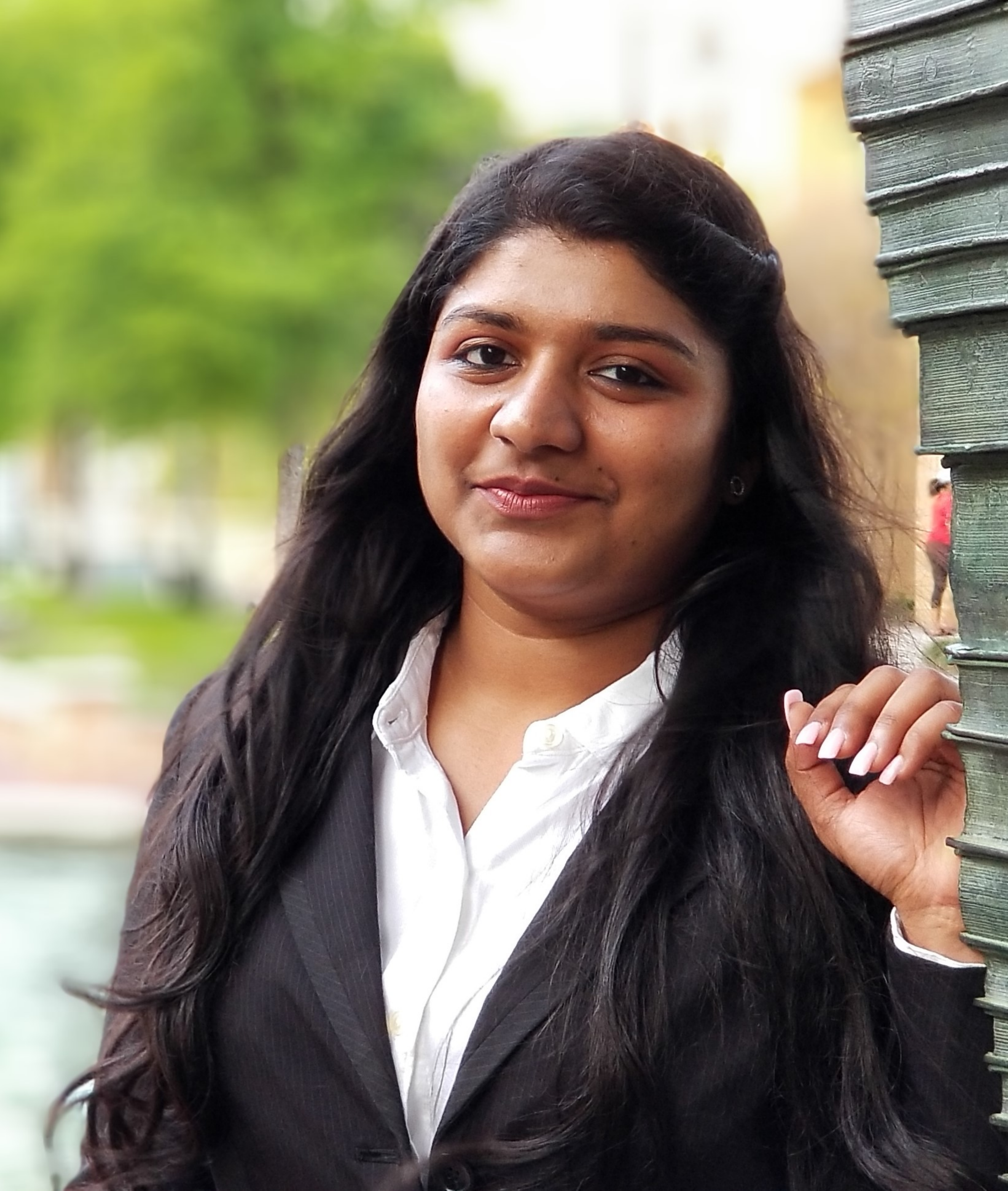}}]  
{Laavanya Rachakonda} (S'18) earned her Bachelors of Technology (B. Tech) in Electronics and Communication from Jawaharlal Nehru Technological University (JNTU), Hyderabad, India, in 2015. She is currently Ph.D. candidate in the research group at Smart Electronics Systems Laboratory (SESL) at Computer Science and Engineering at the University of North Texas, Denton, TX. Her research interests include smart healthcare applications using artificial intelligence, deep learning approaches and application specific architectures for consumer electronic systems based on the IoT.
\end{IEEEbiography}

\vspace{-1.0cm}

\begin{IEEEbiography}
[{\includegraphics[height=1.3in,keepaspectratio]{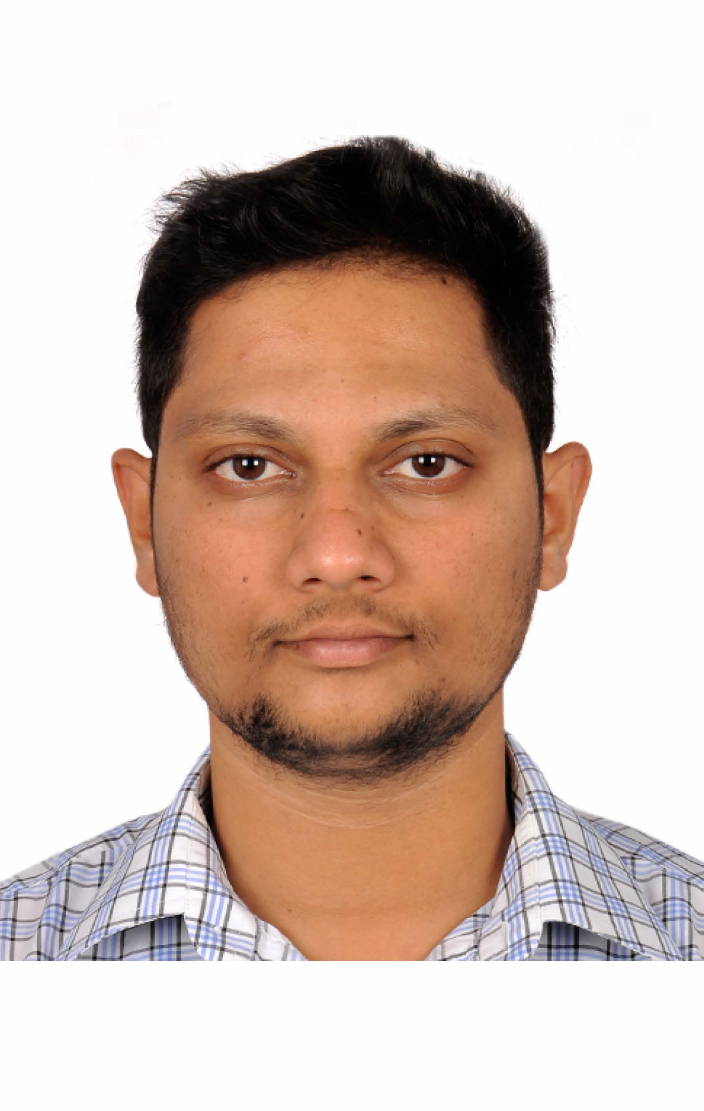}}] 
{Anand K. Bapatla} (S'20) received a Bachelor's of Technology (B. Tech) in Electronics and Communication from Gayatri Vidya Parishad College of Engineering, Visakhapatnam, India, in 2014 and an Master's in Computer Engineering in 2019 from the University of North Texas, Denton, USA. He is currently a Ph.D. candidate in the research group at Smart Electronics Systems Laboratory (SESL) at Computer Science and Engineering at the University of North Texas, Denton, TX. His research interests include smart healthcare and Blockchain applications in Internet of Things (IoT). 
\end{IEEEbiography}

\vspace{-1.0cm}

\begin{IEEEbiography}
[{\includegraphics[height=1.3in,keepaspectratio]{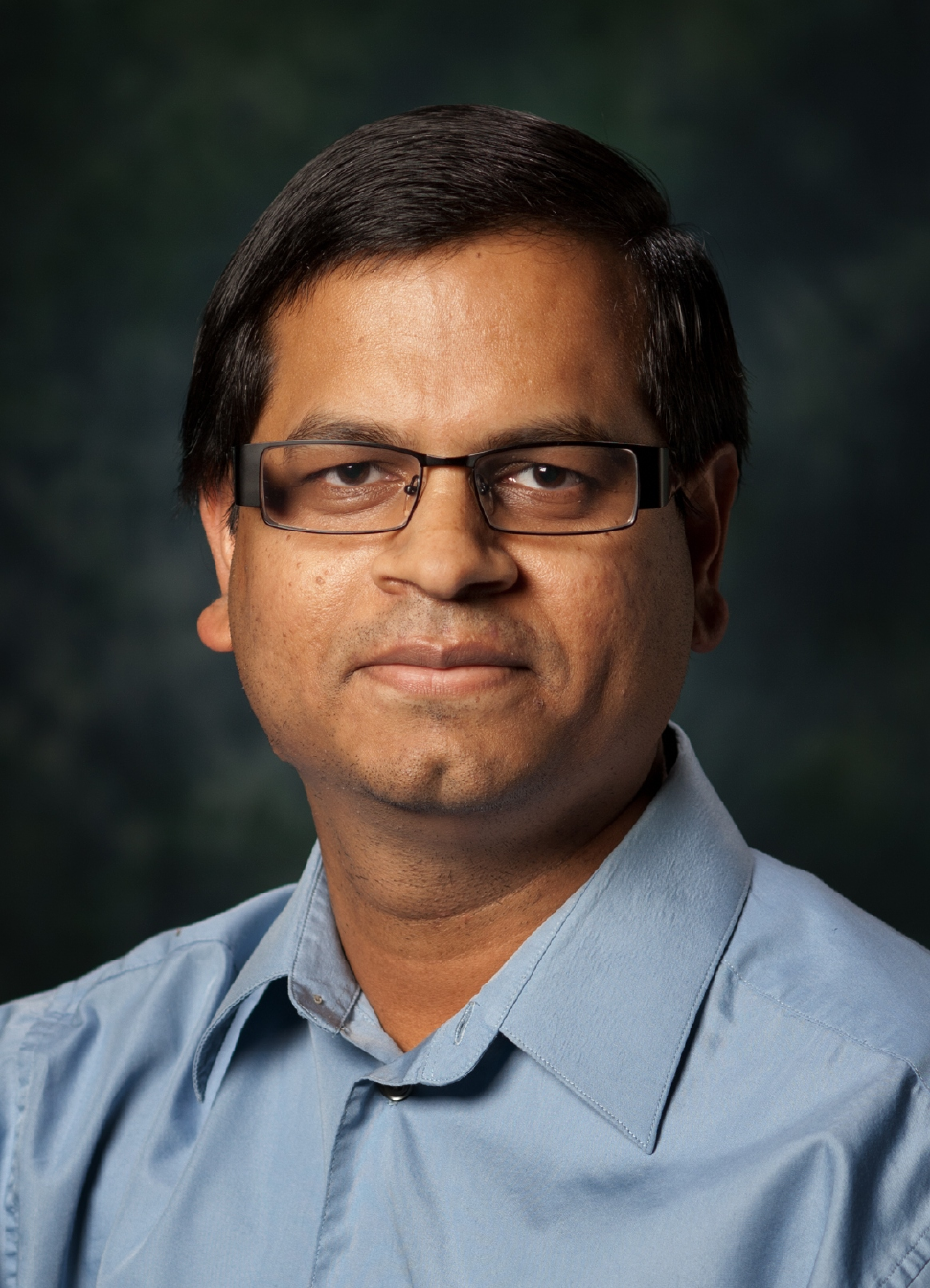}}] 
{Saraju P. Mohanty} (SM'08) received the bachelor's degree (Honors) in electrical engineering from the Orissa University of Agriculture and Technology, Bhubaneswar, in 1995, the master’s degree in Systems Science and Automation from the Indian Institute of Science, Bengaluru, in 1999, and the Ph.D. degree in Computer Science and Engineering from the University of South Florida, Tampa, in 2003. He is a Professor with the University of North Texas. His research is in ``Smart Electronic Systems'' which has been funded by National Science Foundations (NSF), Semiconductor Research Corporation (SRC), U.S. Air Force, IUSSTF, and Mission Innovation. He has authored 350 research articles, 4 books, and invented 4 U.S. patents. His Google Scholar h-index is 36 and i10-index is 133 with 5800+ citations. 
He introduced the Secure Digital Camera (SDC) in 2004 with built-in security features designed using Hardware-Assisted Security (HAS) or Security by Design (SbD) principle. He is widely credited as the designer for the first digital watermarking chip in 2004 and first the low-power digital watermarking chip in 2006.
He is a recipient of 12 best paper awards, Fulbright Specialist Award in 2020, IEEE Consumer Electronics Society Outstanding Service Award in 2020, the IEEE-CS-TCVLSI Distinguished Leadership Award in 2018, and the PROSE Award for Best Textbook in Physical Sciences and Mathematics category in 2016. He has delivered 9 keynotes and served on 5 panels at various International Conferences. 
He has been serving on the editorial board of several peer-reviewed international journals, including IEEE Transactions on Consumer Electronics (TCE), and IEEE Transactions on Big Data (TBD). 
He is the Editor-in-Chief (EiC) of the IEEE Consumer Electronics Magazine (MCE). 
He has been serving on the Board of Governors (BoG) of the IEEE Consumer Electronics Society, and has served as the Chair of Technical Committee on Very Large Scale Integration (TCVLSI), IEEE Computer Society (IEEE-CS) during 2014-2018. 
He has mentored 2 post-doctoral researchers, and supervised 10 Ph.D. dissertations, 26 M.S. theses, and 10 undergraduate projects.
\end{IEEEbiography}

\vspace{-1.0cm}

\begin{IEEEbiography}
[{\includegraphics[height=1.3in,keepaspectratio]{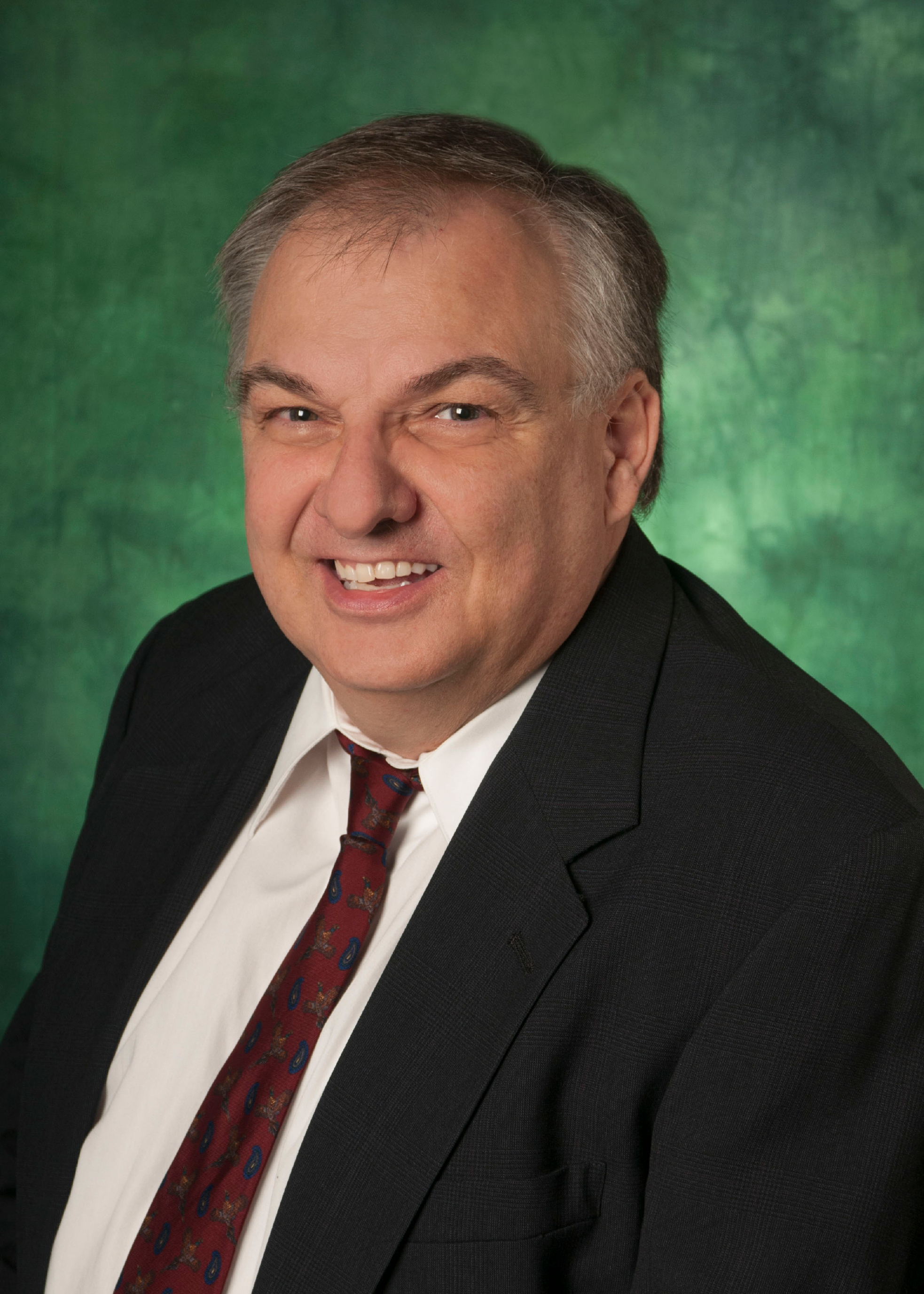}}]
{Elias Kougianos} (SM'07) received a BSEE from the University of Patras, Greece in 1985 and an MSEE in 1987, an MS in Physics in 1988 and a Ph.D. in EE in 1997, all from Louisiana State University. 
From 1988 through 1998 he was with Texas Instruments, Inc., in Houston and Dallas, TX. 
In 1998 he joined Avant! Corp. (now Synopsys) in Phoenix, AZ as a Senior Applications engineer and in 2000 he joined Cadence Design Systems, Inc., in Dallas, TX as a Senior Architect in Analog/Mixed-Signal Custom IC design. He has been at UNT since 2004. He is a Professor in the Department of Electrical Engineering, at the University of North Texas (UNT), Denton, TX. His research interests are in the area of Analog/Mixed-Signal/RF IC design and simulation and in the development of VLSI architectures for multimedia applications. 
He is an author of over 140 peer-reviewed journal and conference publications.  
\end{IEEEbiography}

\end{document}